\documentclass[twocolumn]{aastex63}

\usepackage{float,graphicx,amsmath,multirow}
\usepackage{color}
\usepackage[version=4]{mhchem}
\usepackage{booktabs}
\usepackage{lineno}
\usepackage{xcolor}
\usepackage{xifthen}
\usepackage[normalem]{ulem}
\newcommand{\stkout}[1]{\ifmmode\text{\sout{\ensuremath{#1}}}\else\sout{#1}\fi}
\newcommand{\edited}[2]{\ifthenelse{\isempty{#1}}{\textbf{#2}}{\ifthenelse{\isempty{#2}}{\textcolor{gray}{\stkout{#1}}}{\textcolor{gray}{\stkout{#1}} \textcolor{red}{#2}}}}

\usepackage{xstring,xifthen}
\newcommand{\replacedecimal}[2]{\ifthenelse{\isin{.}{#1}}{\text{\StrBefore{#1}{.}}\ensuremath{\overset{#2}{.}}\text{\StrBehind{#1}{.}}}{#1\ensuremath{^{#2}}}}
\newcommand{\hms}[3]{\ensuremath{#1\overset{\text{h}}{\phantom{.}}#2\overset{\text{m}}{\phantom{.}}\replacedecimal{#3}{\text{s}}}}
\newcommand{\dms}[3]{\ensuremath{#1\overset{\circ}{\phantom{.}}#2\overset{\prime}{\phantom{.}}\replacedecimal{#3}{\prime\prime}}}
\newlength{\VSpaceBeforeTabBib}
\newlength{\VSpaceBeforeTabFoot}
\newcommand*\tablefootname{Notes}
\newcommand*\tablefootfont{\small}
\newcommand*\tablefootnamefont{\small\bfseries}
\newcommand\tablefoot[1]{\VSpaceBeforeTabBib=1ex%
  \par\vspace{\VSpaceBeforeTabFoot}
  \noindent
  \begin{minipage}{\linewidth}
    {\tablefootnamefont\tablefootname.}~%
    \tablefootfont
    \ignorespaces
    #1%
  \end{minipage}%
}
\newcommand*\tablefootmark[1]{%
  \unskip
  \hbox{\textsuperscript{\normalfont\itshape\ignorespaces#1}}%
  \,%
  \ignorespaces
}
\newcommand\tablefoottext[2]{%
  \hbox{\textsuperscript{\normalfont({\itshape\ignorespaces#1})}}%
  ~%
  \ignorespaces
  #2\ \ignorespaces%
}

\begin{document}

\title{Maser Activity of Organic Molecules toward Sgr B2(N)}
\author[0000-0003-2760-2119]{Ci Xue}
\affiliation{Department of Chemistry, Massachusetts Institute of Technology, Cambridge, MA 02139, USA}
\affiliation{Department of Chemistry, University of Virginia, Charlottesville, VA 22904, USA}

\author[0000-0001-9479-9287]{Anthony Remijan}
\affiliation{National Radio Astronomy Observatory, Charlottesville, VA 22903, USA}

\author[0000-0001-7199-2535]{Alexandre Faure}
\affiliation{Université Grenoble Alpes, CNRS, IPAG, F-38000 Grenoble, France}

\author[0000-0003-3168-5922]{Emmanuel Momjian}
\affiliation{National Radio Astronomy Observatory, Socorro, NM 87801, USA}

\author[0000-0001-6492-0090]{Todd R. Hunter}
\affiliation{National Radio Astronomy Observatory, Charlottesville, VA 22903, USA}

\author[0000-0002-8932-1219]{Ryan A. Loomis}
\affiliation{National Radio Astronomy Observatory, Charlottesville, VA 22903, USA}

\author{Eric Herbst}
\affiliation{Department of Chemistry, University of Virginia, Charlottesville, VA 22904, USA}

\author[0000-0003-1254-4817]{Brett McGuire}
\affiliation{Department of Chemistry, Massachusetts Institute of Technology, Cambridge, MA 02139, USA}
\affiliation{National Radio Astronomy Observatory, Charlottesville, VA 22903, USA}

\correspondingauthor{Ci Xue}
\email{cixue@mit.edu}

\begin{abstract}
At centimeter wavelengths, single-dish observations have suggested that the Sagittarius (Sgr) B2 molecular cloud at the Galactic Center hosts weak maser emission from several organic molecules, including \ce{CH2NH}, \ce{HNCNH}, and \ce{HCOOCH3}. However, the lack of spatial distribution information of these new maser species has prevented us from assessing the excitation conditions of the maser emission as well as their pumping mechanisms. Here, we present a mapping study toward Sgr B2 North (N) to locate the region where the complex maser emission originates. We report the first detection of the Class~I methanol (\ce{CH3OH}) maser at 84 GHz and the first interferometric map of the methanimine (\ce{CH2NH}) maser at 5.29\,GHz toward this region. In addition, we present a tool for modeling and fitting the unsaturated molecular maser signals with non-LTE radiative transfer models and Bayesian analysis using the Markov-Chain Monte Carlo approach. These enable us to quantitatively assess the observed spectral profiles. The results suggest a two-chain-clump model for explaining the intense \ce{CH3OH} Class~I maser emission toward a region with low continuum background radiation. By comparing the spatial origin and extent of maser emission from several molecular species, we find that the 5.29\,GHz \ce{CH2NH} maser has a close spatial relationship with the $84\,\mathrm{GHz}$ \ce{CH3OH} Class~I masers. This relationship serves as observational evidence to suggest a similar collisional pumping mechanism for these maser transitions. 
\end{abstract}


\section{Introduction \label{sec:maser-intro}}
Maser emission from molecular species has provided critical insights into the kinetic properties of young stars and has even enabled monitoring of their accretion processes \citep[][and references therein]{2018ASPC..517..321H}. Energetic events (e.g., outflows and collimated jets) and infrared radiation surrounding massive protostars provide the energy to overpopulate the upper energy levels of specific molecular transitions and amplify the molecular emission intensities \citep{2016A&A...585A..71M, 2016A&A...596L...2S}. Common species exhibiting maser emission include, but are not limited to, \ce{OH}, \ce{SiO}, \ce{H2O}, \ce{H2CO}, \ce{NH3}, and \ce{CH3OH}. An empirical taxonomy based on pumping mechanisms classifies molecular masers into two categories \citep{1987Natur.326...49B, 2009ApJ...702.1615C}; \ce{H2O} and Class~I \ce{CH3OH} masers are most often excited via collisions and tend to reside in shocked material associated with outflows and expanding \ion{H}{2} regions \citep[e.g.][]{1974A&A....36..217B, 1989ApJ...347L..35E, 2014MNRAS.439.2584V, 2017ApJS..230...22T}, while \ce{OH} and Class~II \ce{CH3OH} masers are excited via radiation and are generally close to young stellar objects and luminous infrared sources \citep[e.g.,][]{1994A&A...291..569S, 1997A&A...324..211S, 2002MNRAS.331..521C, 2005MNRAS.360..533C}. The most extensively studied Class I \ce{CH3OH} masers include the $4_{-1}-3_{0}\ E$ transition at $36\,\mathrm{GHz}$ and the $7_{0}-6_{1}\ A$ transition at $44\,\mathrm{GHz}$, while Class II \ce{CH3OH} masers include the $5_{1}-6_{0}\ A$ transition at $6.7\,\mathrm{GHz}$ and the $2_{0}-3_{-1}\ E$ transition at $12.2\,\mathrm{GHz}$ \citep[e.g.][]{2014MNRAS.439.2584V,2022ApJS..258...19S}.

By correlating the maser emission from various species with a sample of star-forming regions at different stages, statistical analyses have also suggested that masers may help to trace the evolutionary phases of massive stars \citep[e.g. Figure 6 in][]{Breen2010}. For example, \citet{1989A&A...213..339F} found that isolated \ce{H2O} masers tend to appear in sources at earlier evolutionary stages than \ce{OH} masers based on a survey toward 74 star-forming regions. In addition, Class~II \ce{CH3OH} masers, which appear exclusively in massive star-forming regions \citep{Breen2013,Xu2008}, typically have a significant overlapping timeline with \ce{H2O} masers and arise before the \ce{OH} masers and ultra-compact (UC) \ion{H}{2} regions turn on \citep{2003A&A...410..597W,2005A&A...434..613S}.

However, within the picture of massive star formation, the position for Class~I \ce{CH3OH} masers in an evolutionary sequence is less clear. Initial comparisons of the mid-infrared properties of the sample of \ce{CH3OH} maser sources suggested that Class~I masers are signposts of the earliest stage of massive star formation, even prior to \ce{H2O} masers \citep{2006ApJ...638..241E}. In contrast, detailed mapping of multiple Class~I \ce{CH3OH} maser transitions toward a luminous young stellar object, IRAS\,16547-4247, led \citet{2006MNRAS.373..411V} to argue that Class~I \ce{CH3OH} masers can appear earlier than the Class~II masers but also last long enough to coexist with the \ce{OH} masers. Subsequent observations toward the massive protocluster G18.67+0.03 further suggested that the Class~I \ce{CH3OH} masers might associate with both young sites and relatively evolved UC \ion{H}{2} regions within the same protocluster \citep{2012ApJ...760L..20C}. In any case, high angular resolution imaging is essential for determining which protostar(s) within a protocluster currently power these masers \citep{2017ApJS..230...22T}.

Recently, new maser species, such as \ce{HNCO} and \ce{CH2NH}, have been discovered to exhibit compelling roles in inferring fine-scale structures and gas dynamics in the maser region. In particular, the \ce{HNCO} maser was found to be associated with accretion flows, which helped to build an arm-spiral-structure model of one particular high-mass young stellar object \citep{2020NatAs...4.1170C}; and the \ce{CH2NH} maser toward galactic nuclei helps to probe the inner nuclear processes in extragalactic environments \citep{2021A&A...654A.110G}. The pumping mechanism of these new maser species, however, is less constrained, as is their relation to other common maser species. Observations of these species in maser-active regions that contain both known collisionally and radiatively pumped masers are therefore critical as a first step in understanding their pumping mechanisms.

The high-mass star-forming region Sgr B2 North (N) displays rich maser activity from a diverse set of molecules, ranging from common maser species to rare complex molecules \citep{1992PASJ...44..373M, 2004ApJS..155..577M, 2007ApJ...654..971H, 2018ApJS..239...15Q, 2014PASJ...66...31F}. The single-dish PRebiotic Interstellar MOlecular Survey (PRIMOS) project has identified numerous species exhibiting maser activity at centimeter wavelengths. In fact, the discovery of interstellar carbodiimide (\ce{HNCNH}) was only achievable due to its maser emission \citep{2012ApJ...758L..33M}. Radiative transfer calculations have also revealed that most emission lines of both methyl formate (\ce{HCOOCH3}) and \ce{CH2NH} detected in PRIMOS are weak masers amplifying the background continuum radiation of Sgr B2(N) at cm wavelengths \citep{2014ApJ...783...72F, 2018JPCL....9.3199F}. However, the large ${\sim}2.5\arcmin\times2.5\arcmin$ field of view of PRIMOS at C band ($\sim 5$ GHz) includes both the North and Main components of Sgr B2, complicating the interpretation of the spectral profiles. Therefore, it is crucial to have interferometric observations to disentangle in detail the various possible spatial components of maser emission.

In this work, we present a rigorous imaging study of maser emission from \ce{CH3OH} and \ce{CH2NH} toward Sgr B2(N). We describe the interferometric observations at centimeter (cm) and millimeter (mm) wavelengths in Section~\ref{sec:maser-obs}. A description of the methodology employed to model maser emission is described in Section~\ref{sec:maser-analysis}, which includes the implementation of a Bayesian inference of physical characteristics. Section~\ref{sec:maser-results} presents both the spectra and spatial distribution per maser species. Section~\ref{sec:disc} contains discussions on the metastable states of maser transitions (Section~\ref{sec:energylevels}), pumping mechanisms inferred from the spatial correlations (Section~\ref{sec:pumpingmechanism}), derived physical conditions of masing gas clumps (Section~\ref{sec:physicalconditions}), and a caveat on modeling maser intensity (Section~\ref{sec:caveat}).

\section{Observations \label{sec:maser-obs}}
The interferometric data used here were acquired from two data sets involving the Karl G. Jansky Very Large Array (VLA) in C-band (nominal frequency coverage 4-8\,GHz) and the Atacama Large Millimeter/submillimeter Array (ALMA) in Band 3 (nominal frequency coverage 84–116\,GHz).

\paragraph{VLA Observations} The VLA observations were carried out on 2016 May 27 (project code: 16A-076, P.I.: B.~McGuire) with the C-band receiver. The correlator was configured to deliver 24 subbands or spectral windows, with bandwidths of 4, 8, and 16 MHz, to target various spectral lines. Additionally, nine 128 MHz spectral windows were set up to help with calibration, if needed. The results presented here are based on three spectral windows that delivered channel widths between $0.485\,\mathrm{km\,s^{-1}}$ and $2.31\,\mathrm{km\,s^{-1}}$, depending on the corresponding bandwidth and observing frequency. Detailed observational parameters of each spectral window are presented in Table~\ref{tab:maser-16A-076}. The phase center of the observations was targeted toward Sgr B2 (N), with the field center at $\alpha_\mathrm{J2000}=\hms{17}{47}{19.80}$, $\delta_\mathrm{J2000}=\dms{-28}{22}{17.00}$. The bandpass and absolute flux density scale calibrator was 3C286, while the complex gain calibrator was J1744-3116. The array was in the B configuration with 27 antennas, which covered baselines from $250\,\mathrm{m}$ up to $11\,\mathrm{km}$ and resulted in the maximum recoverable scale of $37\arcsec$ at $4\,\mathrm{GHz}$.

The \texttt{CASA} software package \citep[version 5.6.1,][]{CASA2022} was used for data reduction and imaging analysis of the VLA observations. For each spectral window, the continuum emission was subtracted in the $uv$-plane before imaging the spectral cubes. We used the \texttt{tclean} task with the auto-masking algorithm for imaging with the natural weighting scheme and a cell size of $0.27\arcsec$. The resulting size of the synthesized beam (FWHM) of each image cube has a median value of ${\sim}3.8\arcsec\times1.5\arcsec$. The median of the resultant typical RMS noise levels is ${\sim} 0.9\,\mathrm{mJy\,beam^{-1}}$.

\begin{deluxetable*}{ccrcccc}
    \tablecaption{Summary of VLA Observation Parameters \label{tab:maser-16A-076}}
    \tablehead{
        \colhead{Subband} & \colhead{Frequency Range} & \colhead{Channel} & \colhead{Spectral Resolution}                  & \colhead{BWs}   & \colhead{Synthesized Beam}\tablenotemark{$\ast$} & \colhead{Targeted} \\
        ~             & \colhead{(MHz)}           & \colhead{Number}  & \colhead{(kHz~\textbar~$\mathrm{km\,s^{-1}}$)} & \colhead{(MHz)} & \colhead{$\arcsec \times \arcsec$}          & \colhead{Species}
    }
    \startdata
        5  & 5282.8 -- 5298.8 & 1024 & 15.625~\textbar~0.885 & 16 & $3.351\times1.292$ & \ce{CH2NH}      \\
        17 & 4439.8 -- 4443.8 &  256 & 15.625~\textbar~1.05  &  4 & 3.983$\times$1.542 & \ce{CH2NH}      \\
    \enddata
    \tablecomments{\tablenotetext{\ast}{The synthesized beam size obtained with \texttt{Natural} weighting.}}
\end{deluxetable*}

\paragraph{ALMA Observations} The ALMA observational data were acquired from the ALMA Science Archive of the survey project entitled ``Exploring Molecular Complexity with ALMA (EMoCA)'' (project code: 2011.0.00017.S, P.I.: A.~Belloche). A detailed description of the EMoCA data has been presented in \citet{2016A&A...587A..91B}. In this work, two measurement sets, spectral window 0 of setup S1 and spectral window 0 of setup S4 \citep{2019ApJ...871..112X}, were used for imaging the 84 GHz and 95 GHz \ce{CH3OH} transitions. Different from \citet{2019ApJ...871..112X}, we used the \texttt{tclean} task with the auto-masking algorithm for the imaging analysis, consistent with the VLA imaging process. The resultant synthesized beam size is $2.23\arcsec\times1.67\arcsec$ at $84\,\mathrm{GHz}$ and $1.94\arcsec\times1.49\arcsec$ at $95\,\mathrm{GHz}$ with a median RMS value of $4\,\mathrm{mJy\,beam^{-1}}$.

\section{Analysis \label{sec:maser-analysis}} 
In order to constrain the excitation conditions quantitatively, we used the \texttt{molsim} program to model the molecular spectra with non-LTE radiative transfer models and to explore the physical parameter space with the Markov-Chain Monte Carlo (MCMC) approach \citep{molsim}. 

\subsection{Non-LTE Radiative Transfer Modeling \label{sec:non-lte}}
We implemented the \texttt{RADEX} non-LTE radiative transfer code \citep{2007A&A...468..627V} into \texttt{molsim} written in Python. The module was accelerated with \texttt{Numba} to provide the high performance required for MCMC analysis.

\texttt{RADEX} considers a gas clump consisting of a targeted molecular species and its collision partners, such as \ce{H2} and \ce{e^-}. The gas clump is embedded in a local background radiation field and has a certain geometry (e.g., slab, uniform sphere, expanding sphere). With this physical picture, \texttt{RADEX} iteratively solves for the level population by incorporating an equilibrium with inelastic collisions and the local radiation field. Lastly, it computes the peak optical depth, excitation temperature and peak intensity of each line based on the level population and the line properties, assuming the local radiation field as the only source of background continuum emission. The model can be described with physical parameters including kinetic temperatures ($T_\mathrm{k}$), densities of collision partners ($n_\mathrm{c}$), background radiation field temperatures ($T_\mathrm{bg}$), source velocity ($V_\mathrm{lsr}$), line widths ($dV$), and molecular column densities ($N_\mathrm{col}$).

In our implementation, we construct a modification and extend it to support a continuum temperature ($T_\mathrm{cont}$) different from the local radiation field temperature ($T_\mathrm{rad}$\footnote{\footnotesize{In \texttt{RADEX}, the local radiation temperature is denoted as $T_\mathrm{bg}$.}}) and allow multiple components aligned perfectly along the line of sight (LOS). In the case of a population inversion, $T_\mathrm{cont}$ serves as background radiation that stimulates the maser gain and is amplified by the foreground gas clump. We assume the continuum source and the gas components to be spatially distant from each other such that the continuum radiation is not included in the excitation analysis and the embedded radiation field of each component is only dominated by the local radiation field. Detailed justifications for this assumption will be discussed in Section~\ref{sec:physicalconditions}. With this assumption, the level population of each component is computed with a distinct excitation condition using \texttt{RADEX}. The frequency-dependent optical depth and source function of the $i$-th component are reconstructed by summing over all lines as follows:
\begin{equation}
    \tau_{i}(\nu) = \sum_{j} \tau^\mathrm{peak}_{i, j} \phi_{i}(\nu; \nu_{0, j}),
\end{equation}
\begin{equation}
    S_{i}(\nu) = \frac{\sum_{j} \tau^\mathrm{peak}_{i, j} \phi_{i}(\nu; \nu_{0, j}) B_{\nu}(T_{\mathrm{ex}, i, j})}{\tau_{i}},
\end{equation}
\begin{equation}
    \phi_{i}(\nu; \nu_0) = G\left(\nu; \nu_0 \left(1 - \frac{V_{\mathrm{lsr}, i}}{c}\right), \frac{dV_{i}}{\sqrt{8\ln2}} \right),
\end{equation}
\begin{equation}
    G(x; \mu, \sigma) = \exp\left(-\frac{(x - \mu)^2}{2\sigma^2}\right).
\end{equation}
Here, $\tau^\mathrm{peak}_{i, j}$ and $T_{\mathrm{ex}, i, j}$ are the optical depth and excitation temperature returned by \texttt{RADEX} for the $j$-th line of the $i$-th component, $B_\nu$ is the Planck function, $\phi_{i}$ is the line profile for the $i$-th component using the Gaussian function $G$, and $\tau_{i}$ and $S_{i}$ are the total optical depth and source function with all lines being accounted for.

Using these definitions, for an $n$-component model where the continuum radiation is first processed by the first component, then the second component and so on, the final intensity with the continuum subtracted can be expressed as
\begin{multline}
    I_{\nu} = B_{\nu}(T_\mathrm{cont}) \times (e^{-\sum_{i=1}^{n} \tau_{i}} - 1)\\
    + \sum_{i=1}^{n} S_{i} \times (1 - e^{-\tau_i}) \times e^{- \sum_{j=i+1}^{n} \tau_j}.
\end{multline}
In the case of two components, the continuum-subtracted intensity is
\begin{multline}
    I_{\nu} = B_{\nu}(T_\mathrm{cont}) \times (e^{-\tau_1-\tau_2} - 1)\\
    + S_{1} \times (1 - e^{-\tau_1}) \times e^{-\tau_2}\\
    + S_{2} \times (1 - e^{-\tau_2}).
\end{multline}
It is worthwhile to note that the expressions reduce back to that of \texttt{RADEX} in the case of one component when the continuum temperature is equal to the local radiation temperature (i.e. $T_{cont}=T_{rad}=T_{bg}$). The calculations in \texttt{molsim} reproduces the results returned by \texttt{RADEX} down to machine precision when the results are converged.

In this work, we assume that the local radiation is contributed by the cosmic microwave background only, i.e., $T_\mathrm{rad} = 2.725~\mathrm{K}$ \citep{Mather1999} and a uniform spherical geometry. We only consider \ce{H2} as the collision partner, assuming a kinetic temperature-dependent thermal ortho-to-para ratio (OPR), which can be approximated by:
\begin{equation}
    \mathrm{OPR}(T_\mathrm{k}) =
    \begin{cases}
        9\,e^{-170.6~\mathrm{K}/T_\mathrm{k}} & \text{if $T_\mathrm{k} \leq 150~\mathrm{K}$},\\
        3 & \text{if $T_\mathrm{k} > 150~\mathrm{K}$}.
    \end{cases}
\end{equation}
We left all the rest as free parameters and allowed them to be fitted using the MCMC approach as described below.

\subsection{MCMC Fitting \label{sec:mcmc}}
For line fitting, we used a Bayesian approach to compare models with different parameters and calculate their probability distributions. This approach provides an inference by conditioning the data on priors and sampling posterior distributions. In practice, we use the Markov chain Monte Carlo (MCMC) approach to explore parameter space.

We use a logarithmic likelihood distribution with the formula below:
\begin{equation}
    l = \sum_{i} \left[-\ln \left(\sigma_i \sqrt{2\pi}\right) - \frac{1}{2}\frac{(x_i-\mu_i)^2}{\sigma_i^2}\right]
    \label{eqn:log_likelihood}
\end{equation}
where $x$ is the observed spectral line intensity, $\mu$ is the model line intensity, and $\sigma$ is the uncertainty of the observations including an estimated 10\% systematic uncertainty with the standard baseline deviation. The distribution applies to the case where the dataset has a non-uniform uncertainty, in which case we compute the uncertainties for each spectral chunk separately.

\begin{figure*}[p]
    \centering
    \includegraphics[width=0.95\textwidth]{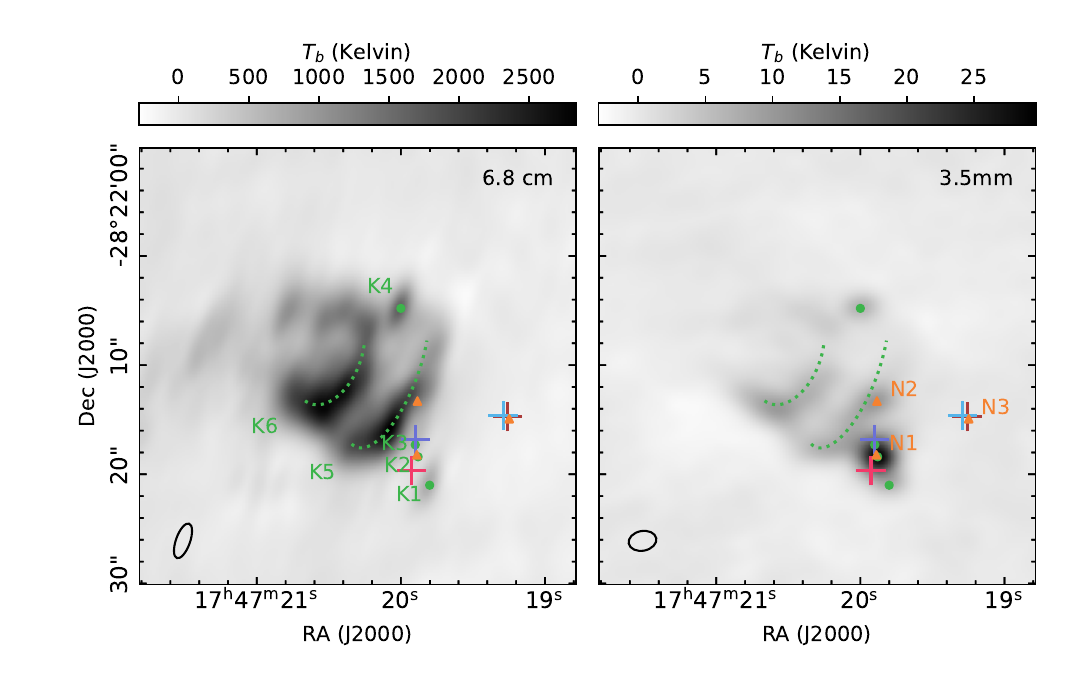}
    \vspace{-6mm}
    \caption{\label{fig:maser-obs-sgrb2}
        Continuum emission toward the Sgr B2(N) region obtained with the VLA at $6.8\,\mathrm{cm}$ and ALMA at $3.5\,\mathrm{mm}$. The cm-wave emission concentrations \citep[K1 -- K6,][]{2015MNRAS.452.3969C} are indicated by green dashed lines and circle markers whereas the molecular hot cores identified in millimeter \citep[N1 -- N3,][]{2017A&A...604A..60B} are indicated by orange triangle markers. The $+$ symbols mark the location of Class~II \ce{CH3OH} masers in blue, the brightest maser spots are marked in rose for the \ce{H2O} maser \citep{2014MNRAS.442.2240W}, in purple for the \ce{OH} maser \citep{2000ApJS..129..159A}, and in firebrick for the \ce{NH3} maser \citep{2022A&A...666L..15Y}.
        }
\end{figure*}

\begin{table*}[p]
    \centering
    \vspace{-3mm}
    \caption{Investigated Maser Transitions \label{tab:maser-transitions-detail}}
    \vspace{-5mm}
    \begin{tabular}{lrlrrrccc}
    \hline
    \hline
        Molecule &Rest Frequency &Transition &$E_\mathrm{u}$ &$S_{ij}\mu^{2}$ &$\log_{10}{\frac{A_{ul}}{\mathrm{s^{-1}}}}$ &$T_\mathrm{B, peak}$\tablefootmark{$\ast$} &$T_\mathrm{ex}$ &$\tau$\\
                 &(MHz) &~ &(K) &($\mathrm{D^2}$) & &(K) &(K)\\
    \midrule
        \ce{CH3OH}  & 84521.172(12) & $5_{-1}-4_{0}\ E$                    & 40.4  & 12.332 &-5.6979& $389.2\pm0.2$&-5.8\tablefootmark{$\dagger$}&-2.44\tablefootmark{$\dagger$}\\
        \ce{CH2NH}  & 5288.962(3)   & $    {1_{1,0}-1_{1,1},}\ F=0-1$      & 11.1  & 0.897  &-8.8109& $\geq 121.8\pm8.6$ &-0.62\tablefootmark{$\ddagger$}&-0.008\tablefootmark{$\ddagger$}\\
        ~           & 5289.712(3)   & $\phm{1_{1,0}-1_{1,1},}\ F=1-0$      & ~     & 0.897  &-9.2881& ~             &-0.54\tablefootmark{$\ddagger$}&-0.009\tablefootmark{$\ddagger$}\\
        ~           & 5289.815(3)   & $\phm{1_{1,0}-1_{1,1},}\ F=2-2$      & ~     & 3.365  &-8.9357& ~             &-0.65\tablefootmark{$\ddagger$}&-0.030\tablefootmark{$\ddagger$}\\
        ~           & 5290.671(3)   & $\phm{1_{1,0}-1_{1,1},}\ F=2-1$      & ~     & 1.121  &-9.4125& ~             &-0.55\tablefootmark{$\ddagger$}&-0.011\tablefootmark{$\ddagger$}\\
        ~           & 5290.828(3)   & $\phm{1_{1,0}-1_{1,1},}\ F=1-2$      & ~     & 1.122  &-9.1908& ~             &-0.75\tablefootmark{$\ddagger$}&-0.008\tablefootmark{$\ddagger$}\\
        ~           & 5291.682(3)   & $\phm{1_{1,0}-1_{1,1},}\ F=1-1$      & ~     & 0.673  &-9.4124& ~             &-0.62\tablefootmark{$\ddagger$}&-0.006\tablefootmark{$\ddagger$}\\
    \hline
    \end{tabular}
    \tablefoot{
    \tablefoottext{$\ast$}{Peak brightness temperature measured at the brightest emission spot for each molecule. The brightest emission spot for 84 GHz \ce{CH3OH} Class~I maser is located near the K4 core at a $V_\mathrm{lsr}$ of 58~$\mathrm{km\,s^{-1}}$. The brightest emission spot for 5.29 GHz \ce{CH2NH} maser is located in the K6 \ion{H}{2} region at a $V_\mathrm{lsr}$ of 83~$\mathrm{km\,s^{-1}}$. The measured $T_\mathrm{B}$ of \ce{CH2NH} represents a lower limit to the $T_\mathrm{B, peak}$ because of the limited angular resolution of the VLA observations.}
    \tablefoottext{$\dagger$}{Characteristics of the \ce{CH3OH} foreground masing clump toward MS1.}
    \tablefoottext{$\ddagger$}{Characteristics of the \ce{CH2NH} masing gas toward MS2.}\\
    }
\end{table*}

\begin{figure*}[p]
    \centering
    \includegraphics[width=0.9\textwidth]{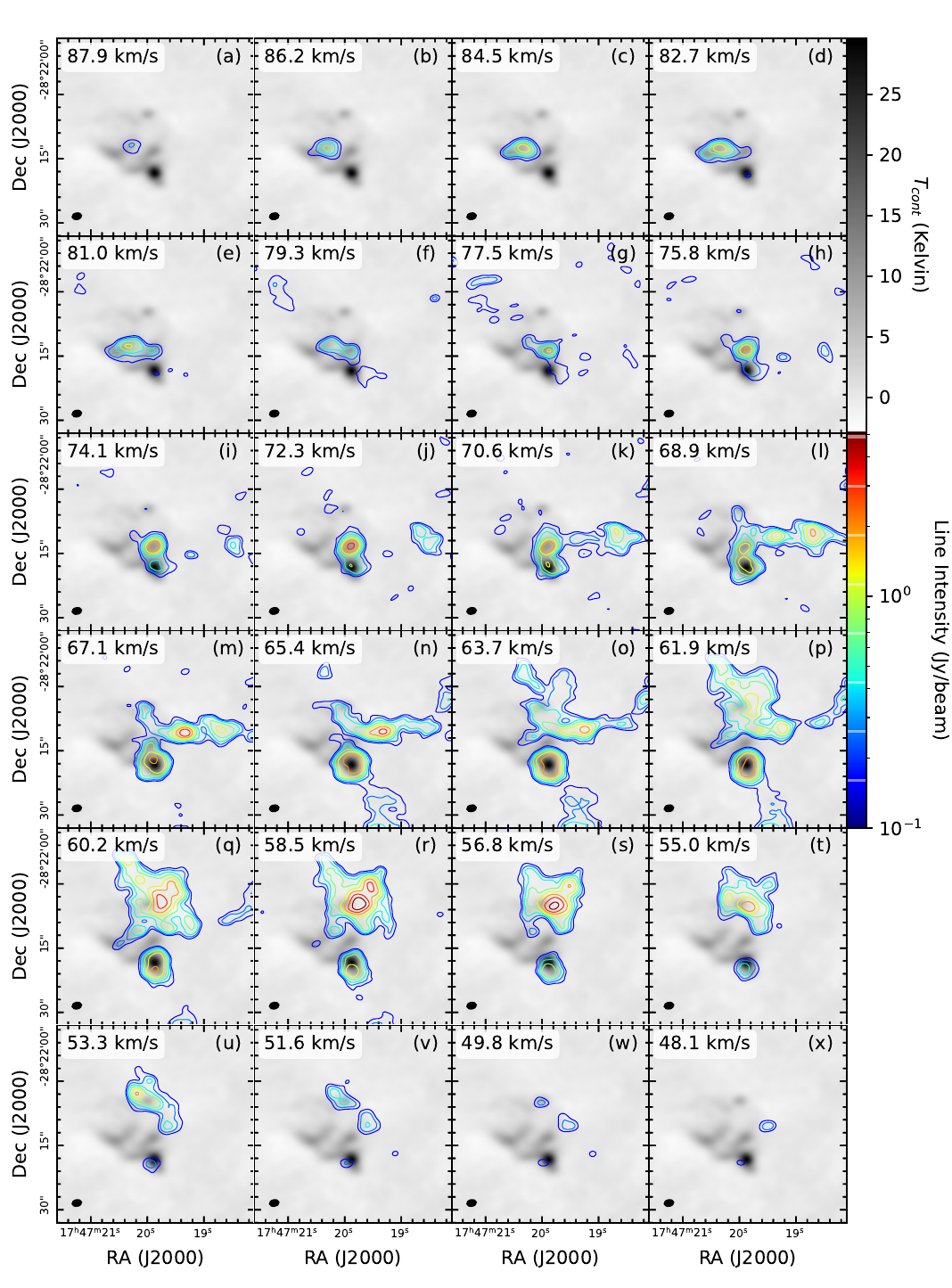}
    \caption{\label{fig:ch3oh-84GHz-map}
        Channel maps of the Class~I \ce{CH3OH} maser at $84\,\mathrm{GHz}$ overlaid on the continuum emission observed at $3.5\,\mathrm{mm}$. The contours levels are $4.0\,\mathrm{mJy\,beam}^{-1} (1\sigma) \times (40$, $65$, $106$, $173$, $282$, $459$, $749$, $1220)$.  The peak value is $8.45\,\mathrm{Jy\,beam^{-1}}$.
    }
\end{figure*}

\begin{figure*}[p]
\begin{minipage}[c][\textheight][c]{\textwidth}
    \centering
    \includegraphics[width=\textwidth]{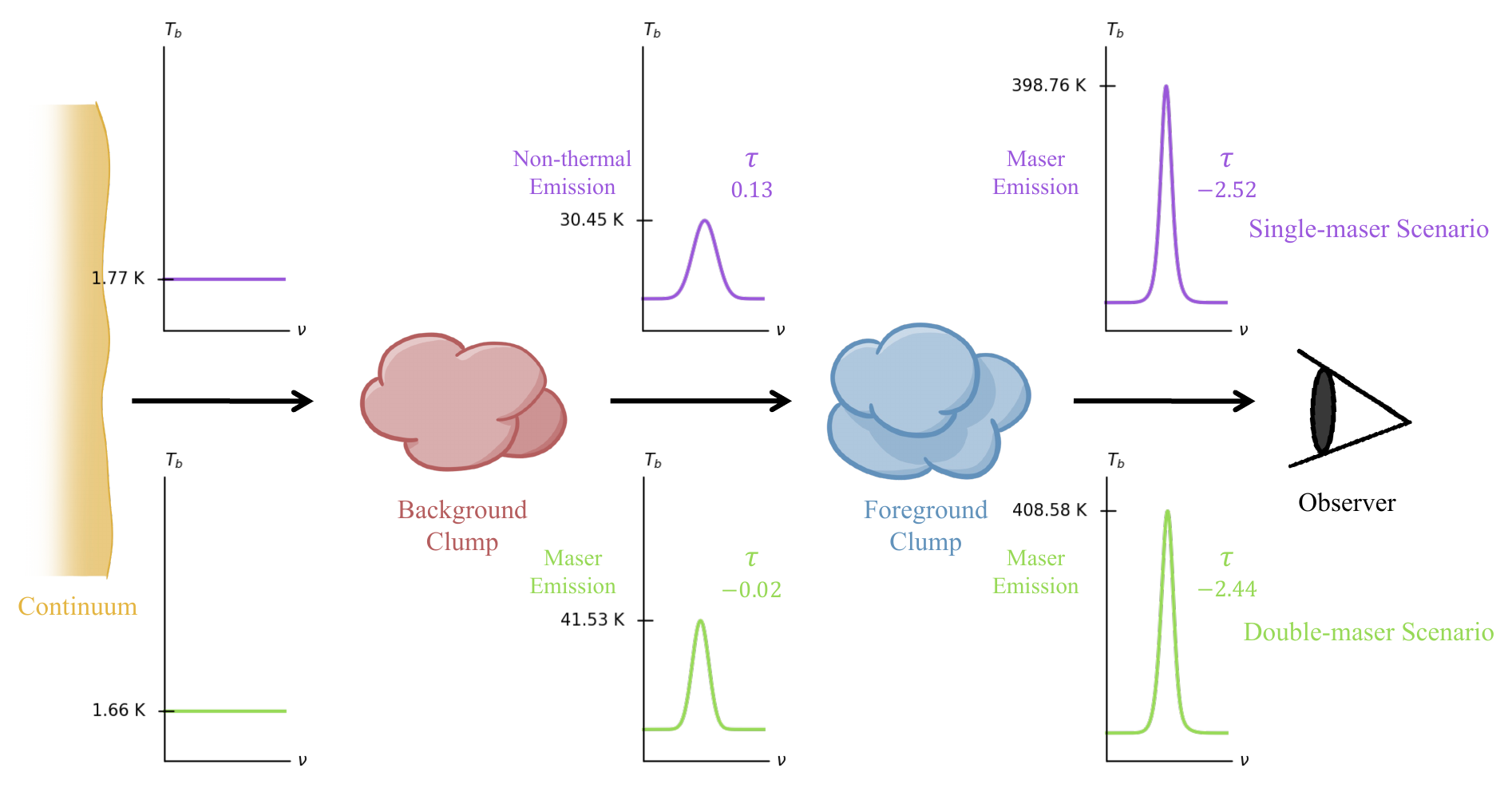}
    \caption{\label{fig:schematic}
        Schematic showing two scenarios of the two-chain-clump model for the 84~GHz \ce{CH3OH} maser, where relatively weak continuum radiation passes through two velocity-coherent gas clumps along the LOS resulting in accumulated maser gains. In the double-maser scenario, the continuum radiation is initially amplified by the background masing clump, followed by further amplification by the foreground masing clump. In the single-maser scenario, however, the foreground masing clump amplifies the non-inverted line emission from the background clump. The peak brightness temperature ($T_b$) and the corresponding optical depth ($\tau$) are annotated in each spectrum.
    }
\end{minipage}
\end{figure*}

The resulting corner plots are shown in Appendix~\ref{apx:mcmc}. A corner plot shows the probability distribution for each pair of parameters and for each parameter individually. In the following, the fitted results are represented by 50$^{th}$, 16$^{th}$, and 84$^{th}$ percentile as the representation value and the range of uncertainty, also known as the 68\% confidence interval. This choice is equivalent to choosing the mean and $+/-1\sigma$ for a normal distribution, which is the case for most of our posterior distributions. One of the advantages of the MCMC technique over a traditional least-squares fit approach is that a larger parameter space can be explored more efficiently. Rather than only returning the parameters that maximize the likelihood function (e.g. least-squares fitting\footnote{\footnotesize{When the posteriors deviate from a Gaussian distribution, the parameters exhibiting the highest likelihood may not necessarily correspond to the optimal parameters.}}), it returns the actual probability of the parameters, which can be used to assess whether parameters are highly co-variant with one another. The correlation between parameters manifests itself as non-separable distributions in the corner plots.

\section{Results \label{sec:maser-results}}
We investigate the maser activities of two transitions in this work. The detailed spectroscopic parameters for each maser transition along with the observational results of the corresponding emission features are summarized in Table~\ref{tab:maser-transitions-detail}. For \ce{CH3OH}, we examined the molecular spectra extracted toward two distinct regions, maser spot 1 (MS1, $\alpha_\mathrm{J2000}=\hms{17}{47}{19.730}$, $\delta_\mathrm{J2000}=\dms{-28}{22}{04.580}$) and maser spot 2 (MS2, $\alpha_\mathrm{J2000}=\hms{17}{47}{20.393}$, $\delta_\mathrm{J2000}=\dms{-28}{22}{12.410}$). Toward both regions, we performed MCMC analyses with a two-chained-component model, as presented in Section~\ref{subsec:ch3oh-ms1} for MS1 and \ref{subsec:ch3oh-ms2} for MS2 respectively. For \ce{CH2NH}, as there is no detectable signal toward MS1, we examined its spectrum extracted toward MS2 only and performed the MCMC analysis with a one-component model as presented in Section~\ref{sec:ch2nh}. 

Empirically, a brightness temperature ($T_\mathrm{B}$) of an emission line much greater than the source physical (kinetic) temperature is considered as an indication of a maser emission rather than a thermally-excited line \citep{2006MNRAS.373..411V, 2011eab..book.....G}. As elaborated in the following sections, our MCMC analyses suggest the $T_k$ for the masing clumps in Sgr B2(N) to be ${\sim}29-37\,\mathrm{K}$. In contrast, as summarized in Table~\ref{tab:maser-transitions-detail}, the observed $T_\mathrm{B}$ values of the molecular emission investigated in this work ($>120$\,K) are significantly higher than the physical temperatures. Furthermore, we also performed a spectral analysis assuming a single excitation temperature to equal $T_k$ for comparison and found that the observed spectral profiles deviate substantially from the thermal model. Additional evidence for maser activities, including the non-Gaussian shape of line profiles, the compact emitting regions, the negative opacity values, and the metastable upper-energy levels, will be discussed more in the following sections. All combine to support the observed emission of \ce{CH3OH} at 84 GHz toward both MS1 and MS2 and \ce{CH2NH} at 5.29 GHz toward MS2 to consist of astronomical (weak) masers.

\begin{figure*}
    \centering
    \includegraphics[width=\textwidth]{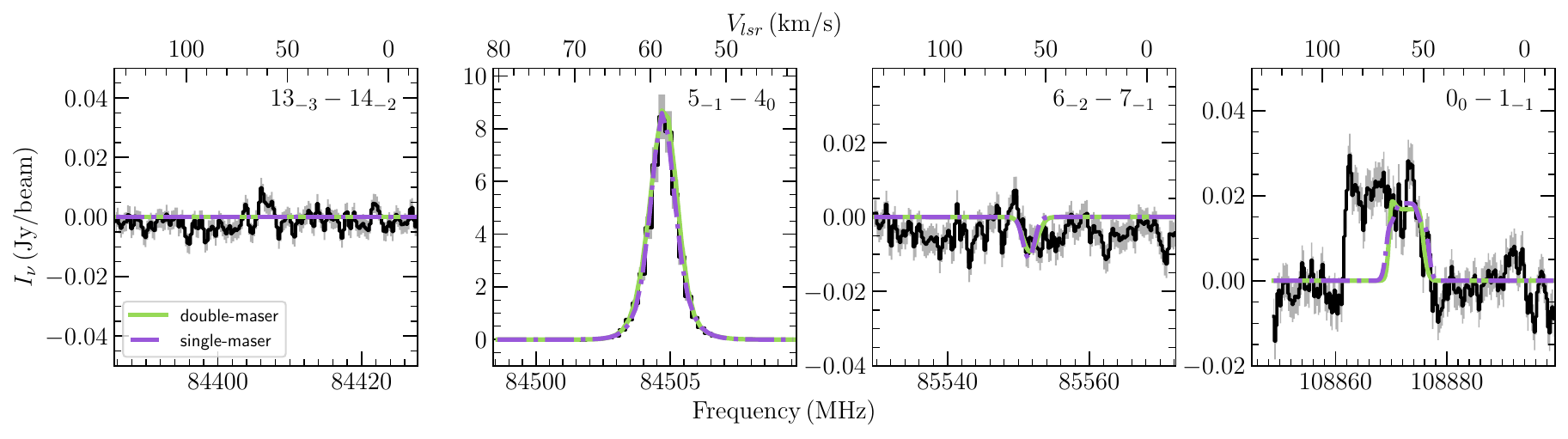}
    \caption{\label{fig:ch3oh-58kms-spectra}
        Observed spectra of the $E$-type \ce{CH3OH} transitions extracted toward MS1 are shown in black with noise levels shaded in gray. Quantum numbers are given at the top right of each panel. The synthetic spectra are overlaid in colors for different scenarios, with the green trace representing the double-maser scenario and the purple trace representing the single-maser scenario. See Table~\ref{tab:MeOH-58kms-par}.
    }
\end{figure*}

\begin{table*}
\centering
\caption{Summary Statistics of the Marginalized Posterior toward MS1 \label{tab:MeOH-58kms-par}}
\begin{tabular}{ccccccc}
    \hline
    \hline
Two-chain-clump Model&$T_{k}$  &$n_{c}$\tablefootmark{$\ast$}   &$V_\mathrm{lsr}$\tablefootmark{$\dagger$} &$dV$    &$N_{col}$ &$T_{cont}$\tablefootmark{$\ddagger$}\\
&			(K)     &(cm$^{-3}$)      &(km s$^{-1}$)  &(km s$^{-1}$)   &(10$^{16}$ cm$^{-2}$)  &(K)\\
\midrule
Single-maser Scenario\tablefootmark{$\mathsection$}&&&&&&$3.4^{+0.1}_{-0.1}$\\
Background &$29.2^{+5.7}_{-5.6}$ &$1.4^{+0.6}_{-0.4}\times10^3$ &$57.9^{+0.1}_{-0.1}$ &$8.1^{+0.2}_{-0.2}$ &$12.1^{+2.6}_{-1.3}$&\\
Foreground &$31.9^{+5.8}_{-4.4}$ &$5.3^{+2.0}_{-1.6}\times10^4$ &$58.5^{+0.1}_{-0.1}$ &$6.2^{+0.2}_{-0.2}$ &$1.3^{+0.2}_{-0.2}$&\\
Double-maser Scenario\tablefootmark{$\mathsection$}&&&&&&$3.3^{+0.2}_{-0.1}$\\
Background &$29.3^{+11.2}_{-8.8}$   &$2.1^{+2.1}_{-0.8}\times10^3$  &$59.4^{+0.2}_{-0.3}$   &$5.8^{+0.2}_{-0.2}$    &$8.1^{+1.5}_{-1.3}$&\\
Foreground &$32.2^{+5.9}_{-4.8}$    &$5.1^{+2.5}_{-2.0}\times10^4$  &$57.3^{+0.2}_{-0.2}$   &$8.0^{+0.3}_{-0.3}$    &$1.5^{+0.4}_{-0.2}$&\\
    \hline
\end{tabular}
    \tablefoot{The quoted uncertainties represent the 16$^{th}$ and 84$^{th}$ percentile ($1\sigma$ for a Gaussian distribution) uncertainties.
    \tablefoottext{$\ast$}{$n_c$ represents the number density of the collision partners, which we take to be $n_{\ce{H2}}$ as hydrogen is assumed to be the dominant partner.}
    \tablefoottext{$\dagger$}{Two velocity components have similar $V_\mathrm{lsr}$ of $\sim58\ \mathrm{km\,s^{-1}}$ along the LOS.}
    \tablefoottext{$\ddagger$}{Continuum temperature for 84-109 GHz band.}
    \tablefoottext{$\mathsection$}{The MCMC analysis revealed two distinct peaks in the posterior distributions, each representing one of the two scenarios. Depending on the scenario, the background clump exhibits maser or non-inverted emission, while the foreground clump always exhibits maser emission.}\\
     }
\end{table*}

As a reference for the relative positions, Figure~\ref{fig:maser-obs-sgrb2} shows the continuum emission from Sgr B2(N) observed at $6.8\,\mathrm{cm}$ (Left) and $3.5\,\mathrm{mm}$ (Right). The free-free continuum emission concentrations at $6.8\,\mathrm{cm}$ include four cores, K1 - K4, and the two bow-shaped \ion{H}{2} regions, K5 and K6 \citep[][marked in green in Figure~\ref{fig:maser-obs-sgrb2}]{2015MNRAS.452.3969C}. At $3.5\,\mathrm{mm}$, three main hot cores, N1 - N3, were revealed by their molecular emission \citep[][marked in orange in Figure~\ref{fig:maser-obs-sgrb2}]{2017A&A...604A..60B}. 

\subsection{Class~I \texorpdfstring{\ce{CH3OH}}{CH3OH} Masers \label{sec:ch3oh}}
Class~I \ce{CH3OH} masers are prevalent in massive star formation regions. In particular, the 44 GHz maser line was observed toward Sgr B2 at $3\arcsec$ resolution \citep{1997ApJ...474..346M}. In contrast, there is limited information about the $5_{-1}-4_{0}\ E$ Class~I \ce{CH3OH} maser at $84\,\mathrm{GHz}$, a member of the same $J+1_{-1}-J_0\ E$-type transition family along with the 36 GHz line \citep[][and references therein]{2019MNRAS.484.5072B}. Interferometric observations of the 84 GHz line in the Milky Way are even rarer \citep{2006MNRAS.373..411V}, as recent work has focused on extragalactic detections \citep{Humire2022,Tiege2018}.

Here, we present the first interferometric ALMA observations of the 84 GHz \ce{CH3OH} masers in a Milky Way object at a resolution of $1.5\arcsec$, as well as the first detection of the 84 GHz maser toward Sgr B2. The spread of 84 GHz maser spots is manifest in the channel images shown in Figure~\ref{fig:ch3oh-84GHz-map} with velocities ranging from $50$ to $86\,\mathrm{km\,s^{-1}}$. The brightest maser spot located ${\sim}3.6\arcsec$ west of the K4 core with a $V_\mathrm{lsr}$ of 58 $\mathrm{km\,s^{-1}}$ (hereafter referred to as MS1) has a peak $T_\mathrm{B}$ of $389.2\pm0.2$ K ($8.454\pm0.004\ \mathrm{Jy\,beam^{-1}}$, Figure~\ref{fig:ch3oh-84GHz-map}~{($q$)-($t$)}). In addition to the brightest spot MS1, another maser knot associated with the K6 \ion{H}{2} region is also identified with a $T_\mathrm{B}$ of $75.4\pm0.4$ K ($1.636\pm0.008\ \mathrm{Jy\,beam^{-1}}$) at a $V_\mathrm{lsr}$ of $83\,\mathrm{km\,s^{-1}}$ (hereafter referred to as MS2), as shown in Figure~\ref{fig:ch3oh-84GHz-map}~($b$)-($e$).

The molecular data file for $E$-type \ce{CH3OH} was taken from the Excitation of Molecules and Atoms for Astrophysics (EMAA) database \footnote{\footnotesize{https://emaa.osug.fr}}, with spectroscopy from the Cologne Database for Molecular Spectroscopy (CDMS) catalog\footnote{\footnotesize{https://www.astro.uni-koeln.de/cdms/catalog}} \citep{2005JMoSt.742..215M} and collision rate coefficients reported by \citet{2010MNRAS.406...95R} for the rotational excitation of \ce{CH3OH} by para- and ortho-\ce{H2} (hereafter p-\ce{H2} and o-\ce{H2}). The calculations of \citet{2010MNRAS.406...95R} were performed for rotational transitions within the three first torsional states of \ce{CH3OH}, $\nu=0, 1, 2$, with a \ce{CH3OH} basis including rotational states $j_1\leq 15$ for p-\ce{H2}($j_2=0$) and $j_1\leq 9$ for o-\ce{H2}($j_2=1$), where $j_1$ and $j_2$ are the rotational quantum numbers. The coupled-states approximation was used to generate rate coefficients at temperatures $10 \leq T_k \leq 200$~K. Because torsionally inelastic transitions were not considered by \citet{2010MNRAS.406...95R}, the EMAA dataset for $E$-\ce{CH3OH} includes rotational transitions within the ground torsional state ($\nu=0$) only. The first excited state ($\nu=1$) of $E$-\ce{CH3OH} opening at 204.2~cm$^{-1}$, the highest level of $E$-\ce{CH3OH} included in the EMAA dataset is the rotational state $9_{-6}$ ($j_1=9, K=-6$) with an energy of 203.9~cm$^{-1}$.

\subsubsection{Masers with Dark Continuum Background -- MS1 \label{subsec:ch3oh-ms1}}
Four uncontaminated $E$-type \ce{CH3OH} transitions covered in the collision data were used to constrain the physical parameters quantitatively using a non-LTE model and MCMC fitting algorithm. The observed spectra extracted toward the brightest maser spot MS1 are shown in Figure~\ref{fig:ch3oh-58kms-spectra}. We first attempted with a single-clump model but found that the observed spectra could not be reproduced with physically realistic parameters even after relaxing the constraints on $T_\mathrm{cont}$ and $T_\mathrm{rad}$. In order to match the bright maser emission at 84.5 GHz, strong background radiation is needed to stimulate the maser gain in the single-clump model but it will also introduce significant absorption features at both 85.5 and 108.9 GHz.

Moreover, as shown in Figure~\ref{fig:ch3oh-84GHz-map}(r), there is no significant mm-wave continuum source at MS1. While this spot also exhibits a strong \ce{CH3OH} emission at 44 GHz \citep[$T_\mathrm{B}$ = 157\,K,][]{1997ApJ...474..346M}, it shows no continuum emission at cm-wavelengths either. The closest continuum source, K4, is offset by $\sim3.6\arcsec$ which corresponds to $\sim$0.14\,pc at the distance of Sgr B2 \citep[8.2 kpc,][]{2019ApJ...885..131R}. In addition, as shown in Appendix~\ref{apx:IR}, there are no point sources associated with this maser spot. The nearest bright infrared source observed with both the Two Micron All Sky Survey (2MASS) and the Galactic Legacy Infrared Midplane Survey Extraordinaire (GLIMPSE) program is located at $\sim5.2\arcsec$ to the west of the maser spot. Instead, the background radiation stimulating the maser gain may be weak and diffuse, causing it to be resolved out by interferometers, such as the free-free and dust emission in the Galactic plane.

As illustrated in Figure~\ref{fig:schematic}, the model we consider consists of two chained gas clumps overlapping in velocity along the LOS, where the foreground clump exhibits maser activity but the background clump can either be also an inverted maser (i.e. the double-maser scenario) or non-inverted emitting (i.e. the single-maser scenario). In both scenarios, this model can account for the strong maser emission with an absence of a bright and compact continuum source. It is analogous to the ``self-amplification" model of two masing clouds that has been used to interpret water megamasers observed in circumnuclear disks in active galactic nuclei \citep{1999ApJ...513..180K}.

A chained two-component model has been constructed to iteratively generate synthetic spectra, each component being described with five parameters, $T_k$, $n_c$, $N_{col}$, $dV$ and $V_\mathrm{lsr}$. We assumed the two components have coherent but not identical values of  $V_\mathrm{lsr}$ of ${\sim}58\ \mathrm{km\,s^{-1}}$ although spatially separated along the LOS. As such, two $V_\mathrm{lsr}$ parameters have the same prior probability distributions but are fit independently. Including both $V_\mathrm{lsr}$ values, the priors for the free parameters and the MCMC analysis are discussed in detail in Appendix~\ref{apx:mcmc}. Posterior probability distributions for each parameter and their covariances were generated with 100 walkers with 50000 samples each\footnote{\footnotesize{Each walker represents a Markov Chain consisting of multiple samples with each sample being randomly sampled within the parameter space.}}.

As revealed in the bimodal posterior distributions, the samples converge toward two distinct peaks, each corresponding to the double-maser and single-maser scenarios, respectively. The statistics of the resulting parameters for each scenario were summarized in Table~\ref{tab:MeOH-58kms-par}, which were used to construct the synthetic spectra for the double-maser scenario shown with the green trace and the single-maser scenario shown with the purple trace in Figure~\ref{fig:ch3oh-58kms-spectra}. Both synthetic profiles well reproduce the observed spectra, capturing both the maser and the optically thick emission features. Our model, unfortunately, falls short in matching the low-frequency side of the emission plateau at 108.87 GHz, corresponding to the $0_{0}-1_{-1}$ transition at a velocity range of $70-90$~$\mathrm{km\,s^{-1}}$. This divergence from the observed spectrum implies the potential presence of another optically-thick velocity component(s) along the LOS. Because neither of the two scenarios completely reproduces the observed data, it is noteworthy that the uncertainties derived from the MCMC results should be considered as a lower limit of the uncertainties.

\begin{figure}
    \centering
    \includegraphics[width=0.48\textwidth]{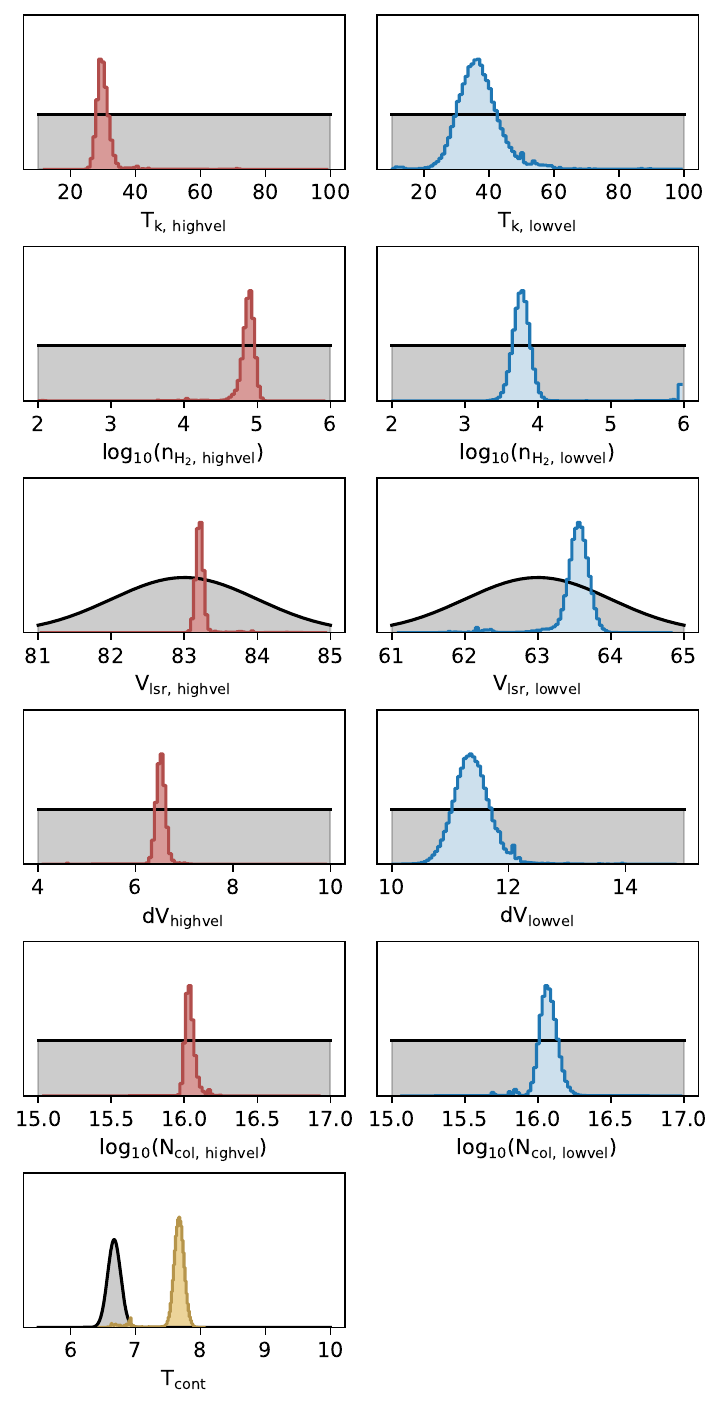}
    \caption{\label{fig:ch3oh-prior-posterior-I} Prior and posterior distributions for the MCMC fit of \ce{CH3OH} $E$-type transitions toward MS2. The prior distribution of each parameter is shown in gray. The posterior distributions for the low-velocity component are shown in blue, whereas those for the high-velocity component are shown in red. The posterior of $T_{cont}$ is shown in yellow. The discrepancy of $T_{cont}$ in the posterior relative to the prior could be be attributed to a certain portion of continuum flux, especially from extended components, being resolved out by ALMA.}
\end{figure}

\begin{figure*}
    \centering
    \includegraphics[width=\textwidth]{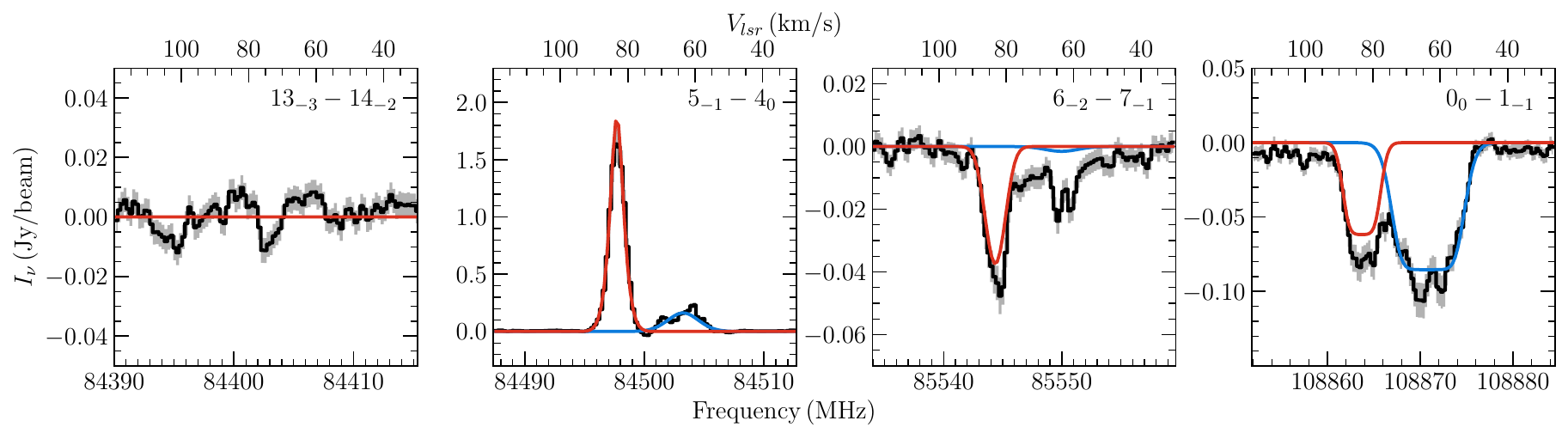}
    \caption{\label{fig:ch3oh-83kms-spectra}
        Observed spectra of the $E$-type \ce{CH3OH} transitions extracted toward MS2 are shown in black with noise levels shaded in gray. Quantum numbers are given at the top right of each panel. The rest frequency for each transition are 84423.769, 84521.172, 85568.131, 108893.945 MHz from left to right. The synthetic spectra are overlaid in colors with the red trace representing the high-velocity component (83~$\mathrm{km\,s^{-1}}$) and the blue trace representing the low-velocity component (64~$\mathrm{km\,s^{-1}}$). See Table~\ref{tab:83kms-par}.
    }
\end{figure*}

\begin{table*}
\centering
\caption{Summary Statistics of the Marginalized Posterior toward MS2 \label{tab:83kms-par}}
\begin{tabular}{ccccccc}
    \hline
    \hline
Molecule &$T_{k}$ &$n_c$    &$V_\mathrm{lsr}$\tablefootmark{$\ast$} &$dV$ &$N_{col}$ &$T_{cont}$\\
&(K) &(cm$^{-3}$) &(km s$^{-1}$) &(km s$^{-1}$) &(cm$^{-2}$) &(K)\\
\midrule
\hspace{0.1em}\vspace{-0.5em}
\ce{CH3OH}  &&&&&&$7.7^{+0.1}_{-0.1}$\tablefootmark{$\dagger$} \\
            &$30.7^{+2.0}_{-1.5}$ &$8.3^{+1.5}_{-1.6}\times10^4$ &$83.3^{+0.1}_{-0.1}$ &$6.6^{+0.1}_{-0.1}$ &$1.1^{+0.1}_{-0.1} \times 10^{16}$&\\
            &$37.3^{+6.4}_{-5.6}$ &$6.6^{+2.1}_{-1.5}\times10^3$ &$63.6^{+0.1}_{-0.2}$ &$11.4^{+0.3}_{-0.3}$ &$1.2^{+0.2}_{-0.2} \times 10^{16}$&\\
\ce{CH2NH}  &&&&&&$2687.6^{+62.4}_{-62.4}$\tablefootmark{$\ddagger$}\\
            &$31.3^{+4.3}_{-4.3}$ &$6.2^{+3.7}_{-2.4}\times10^4$ &$83.5^{+0.3}_{-0.3}$ &$7.7^{+0.5}_{-0.5}$ &$4.8^{+1.1}_{-0.6} \times 10^{14}$&\\
     \hline
     \end{tabular}
	\tablefoot{Similar to Table~\ref{tab:MeOH-58kms-par}.
    \tablefoottext{$\ast$}{Two velocity components are identified along the LOS at a $V_\mathrm{lsr}$ of 83~$\mathrm{km\,s^{-1}}$ and 64~$\mathrm{km\,s^{-1}}$ respectively.}
    \tablefoottext{$\dagger$}{Continuum temperature for 84-109 GHz band.}
    \tablefoottext{$\ddagger$}{Continuum temperature at 5 GHz.}\\
     }
\end{table*}

In the double-maser scenario, the $\tau$ and $T_{ex}$ of the $5_{-1}-4_{0}$ maser transition are found to be $-$2.44 and $-$5.8~K for the foreground clump and $-$0.02 and $-$2027.3~K for the background clump. In the single-maser scenario, the corresponding $\tau$ and $T_{ex}$ values are $-$2.52 and $-$6.5~K for the foreground clump and 0.13 and 245.2~K for the background clump. As an $T_{ex}$ of 245.2~K is significantly higher than the $T_{k}$ value in this case, it is evident that this transition exhibits strong non-inverted emission and is on the verge of a population inversion.

\subsubsection{Masers with Bright Continuum Background -- MS2 \label{subsec:ch3oh-ms2}}
In contrast to MS1, MS2 is aligned with the K6 \ion{H}{2} region along the LOS, which provides background radiation to be amplified by the stimulated emission (Figure~\ref{fig:ch3oh-84GHz-map}d). The spectra extracted toward MS2 are shown in Figure~\ref{fig:ch3oh-83kms-spectra} with two individual velocity components being identified for each of the spectral features. We also performed the spectral analysis for MS2 with a two-component model, where each component has a distinct $V_\mathrm{lsr}$.

The adopted prior and resultant posterior probability distributions for each parameter are depicted respectively in Figure~\ref{fig:ch3oh-prior-posterior-I}. The priors for $V_\mathrm{lsr}$ were chosen to be Gaussian distributions centering on 83~$\mathrm{km\,s^{-1}}$ and 63~$\mathrm{km\,s^{-1}}$ respectively, based on the difference between the observed sky frequency and the rest frequency. Meanwhile, the $T_{cont}$ prior also has a Gaussian distribution, based on the measured continuum level of $6.67\pm0.52$~K. The remaining parameters all have uninformative uniformly distributed priors. Following 100 walkers with 10,000 samples each, all parameters converge to primarily Gaussian-like distributions in our model. The corresponding corner plot of the posterior distributions and the parameter covariances are shown in Appendix~\ref{apx:mcmc}, with the largest covariance observed between \ce{H2} densities and \ce{CH3OH} column densities. This covariance aligns with the fact that \ce{H2} densities and \ce{CH3OH} column densities are related through the maser emission measure, $\xi \propto n_{c} \times N_\mathrm{col}$, which, together with temperature, determines the overall rate coefficient for a collisionally-pumped maser transition \citep{2016A&A...592A..31L}.

The posterior medians from the MCMC inference (Table \ref{tab:83kms-par}) were used to construct the synthetic spectra shown in Figure~\ref{fig:ch3oh-83kms-spectra}. If we take the noise level measured in each passband into account, the synthetic spectral profiles fit satisfactorily with the observed spectra for both the maser and optically thick absorption features. However, the low-velocity component (64~$\mathrm{km\,s^{-1}}$) of the $6_{-2}-7_{-1}$ transition fails to match the observed absorption feature at 85.55 GHz. This absorption feature might be attributed to potential contamination from an unidentified transition. The excitation temperature and opacity of the $5_{-1}-4_{0}$ maser transition were found to be negative, i.e., $-5.6$ K of $T_{ex}$ and $-2.52$ of $\tau$ for the high-velocity component (83~$\mathrm{km\,s^{-1}}$), confirming its maser activity toward MS2.

\subsection{\texorpdfstring{\ce{CH2NH}}{CH2NH} Masers \label{sec:ch2nh}}
\begin{figure*}
    \centering
    \includegraphics[width=\textwidth]{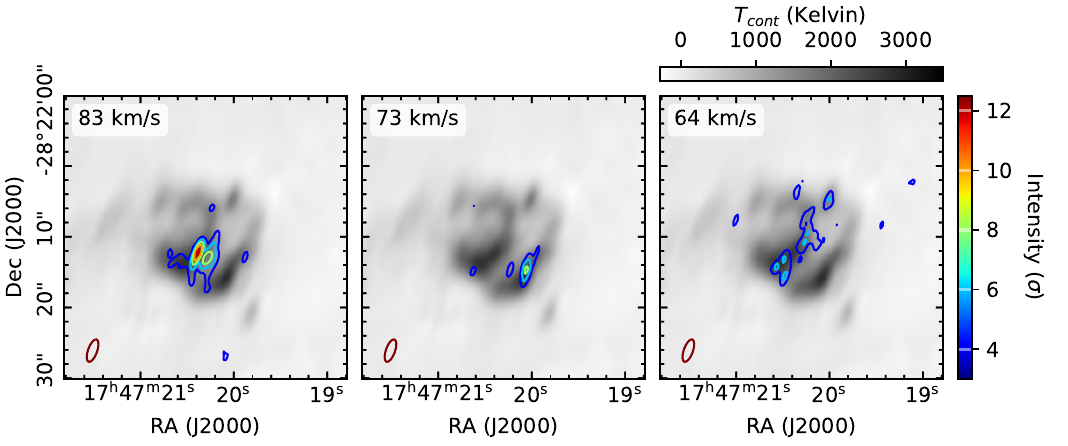}
    \caption{\label{fig:maser-ch2nh-maps}
        Peak intensity maps for each velocity component of the \ce{CH2NH} maser at $5.29\,\mathrm{GHz}$ superimposed on the continuum emission observed at $6.8\,\mathrm{cm}$. The contour levels start at $4\,\sigma$ and continue up to the peak value with a step size of $2\,\sigma$. The values of $\sigma$ and peak values for the 83-, 73-, and 64-$\mathrm{km\,s^{-1}}$ components are 0.80, 0.93, 0.85~$\mathrm{mJy\,beam^{-1}}$, and 12.1, 8.86, 6.60~$\mathrm{mJy\,beam^{-1}}$, respectively.
    }
\end{figure*}

The maser activity of the $1_{1,0}-1_{1,1}$ \ce{CH2NH} transitions at $5.29\,\mathrm{GHz}$ has been revealed toward both the Galactic Center and external galaxies as point source emission \citep{2018JPCL....9.3199F, 2021A&A...654A.110G}. With the VLA observations, we resolve this emission spatially into three velocity components with a $V_\mathrm{lsr}$ of $64$, $73$, and $83\,\mathrm{km\,s^{-1}}$. The peak intensity map for each component is shown in Figure~\ref{fig:maser-ch2nh-maps} respectively. The 83 $\mathrm{km\,s^{-1}}$ and 64 $\mathrm{km\,s^{-1}}$ components are both associated with the K6 \ion{H}{2} region, while the 73-$\mathrm{km\,s^{-1}}$ component is concentrated toward the K5 \ion{H}{2} region. The brightest 83-$\mathrm{km\,s^{-1}}$ component displays a more compact distribution than the other two components, with a $T_\mathrm{B}$ of $121.8\pm8.6\,\mathrm{K}$ ($12.1\pm0.85\ \mathrm{mJy\,beam^{-1}}$). This measured $T_\mathrm{B}$ represents a lower limit to the peak $T_\mathrm{B}$ since the observed emission morphology of the 83-$\mathrm{km\,s^{-1}}$ component may result from point sources convolved with the coarse VLA beam. Nevertheless, this component of the \ce{CH2NH} emission is coherent with that of the 84~GHz \ce{CH3OH} maser emission both in the velocity domain and spatial position. Therefore, the significant brightness, highly compact morphology, and coherence with other maser emission suggest maser activity in the 83-$\mathrm{km\,s^{-1}}$ component.

The molecular data file for \ce{CH2NH} was taken from the EMAA database. Extended from the calculations in \citet{2018JPCL....9.3199F}, we calculated the collisional data for rotational transitions of \ce{CH2NH} among the first 49 rotational levels, i.e. up to the state $8_{2, 6}$ at 99.5~cm$^{-1}$, and for kinetic temperatures in the range 10 -- 150~K. The detail calculations of the collisional rate coefficients are presented in Appendix~\ref{apx:ch2nh-coll}.

We performed the MCMC analysis with a one-component model using the prior probability distributions depicted in Figure~\ref{fig:ch2nh-prior-posterior}. The VLA observations only cover two transitions of \ce{CH2NH} while the other transition at $4.44\,\mathrm{GHz}$ has no signal over the noise levels. As such, the data are insufficient to simultaneously constrain the physical characteristics so the \ce{CH2NH} MCMC analysis requires a confined choice of priors on some parameters. We opted to use $E$-type \ce{CH3OH} discussed in Section~\ref{subsec:ch3oh-ms2} for physical constraints, where the 83~$\mathrm{km\,s^{-1}}$ components between \ce{CH3OH} and \ce{CH2NH} are spatially coherent and are both associated with the K6 H-II region. We therefore extracted the \ce{CH2NH} spectrum toward the same position of MS2. The posteriors on $T_k$ and $n_{\ce{H2}}$ obtained from \ce{CH3OH} (Table~\ref{tab:83kms-par}; Figure~\ref{fig:ch3oh-prior-posterior-I}; Figure~\ref{fig:MeOH_triangle_83kms}) are adopted as MCMC priors for the \ce{CH2NH} model. The priors for $V_\mathrm{lsr}$ and $dV$ are chosen to be Gaussian distributions, informed by the strongest emission feature (12$\sigma$) detected in the data. The Gaussian prior for $T_{cont}$ is based on the continuum level of $2687\pm58$~K measured by the VLA. Meanwhile, the $N_{col}$ of \ce{CH2NH} has a non-informative uniformly distributed prior. 

\begin{figure}
    \centering
    \includegraphics[width=0.45\textwidth]{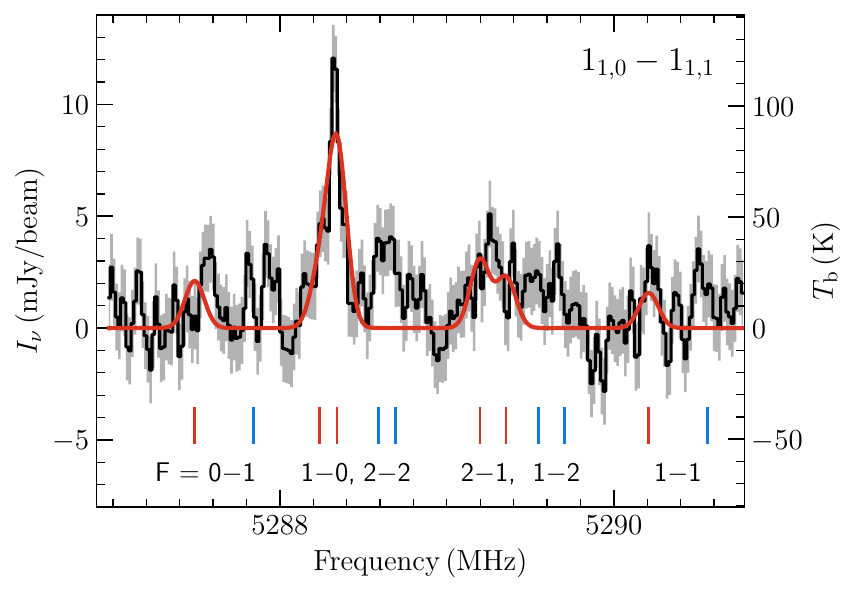}
    \caption{\label{fig:maser-ch2nh-spec} Observed spectrum of the \ce{CH2NH} $1_{1, 0}-1_{1, 1}$ masing transition extracted toward MS2 is shown in black with noise levels shaded in gray. The synthetic spectrum with hyperfine splitting is overlaid in red for the 83-$\mathrm{km\,s^{-1}}$ components. The $F=1-0$ and $F=2-2$ hyperfine components are blended. Each hyperfine component is denoted by a vertical bar, with the red bars corresponding to a $V_\mathrm{lsr}$ of $83\,\mathrm{km\,s^{-1}}$ and the blue bars corresponding to $64\,\mathrm{km\,s^{-1}}$ in blue.
    }
\end{figure}

As shown in the posterior distributions in Figure~\ref{fig:ch2nh-prior-posterior}, following 10,000 samples with each of the 100 walkers, the samplers converge at a $N_{col}$ of $4.76^{+1.09}_{-0.64} \times 10^{14}\ \mathrm{cm^{-2}}$. The synthetic line profile is shown in Figure~\ref{fig:maser-ch2nh-spec}, reproducing six hyperfine components of the $1_{1,0}-1_{1,1}$ transition within the noise level. The six hyperfine components have $T_{ex}$ in the range of -0.75 -- -0.55~K, $\tau$'s in the range of -0.03 -- -0.006, and, in particular, the $F=2-2$ transition has a $T_{ex}$ of -0.65~K and a $\tau$ of -0.03.

\section{Discussion \label{sec:disc}}
\subsection{Metastable Energy Levels of Maser Transitions} \label{sec:energylevels}
\begin{figure*}
    \centering
    \includegraphics[width=\textwidth]{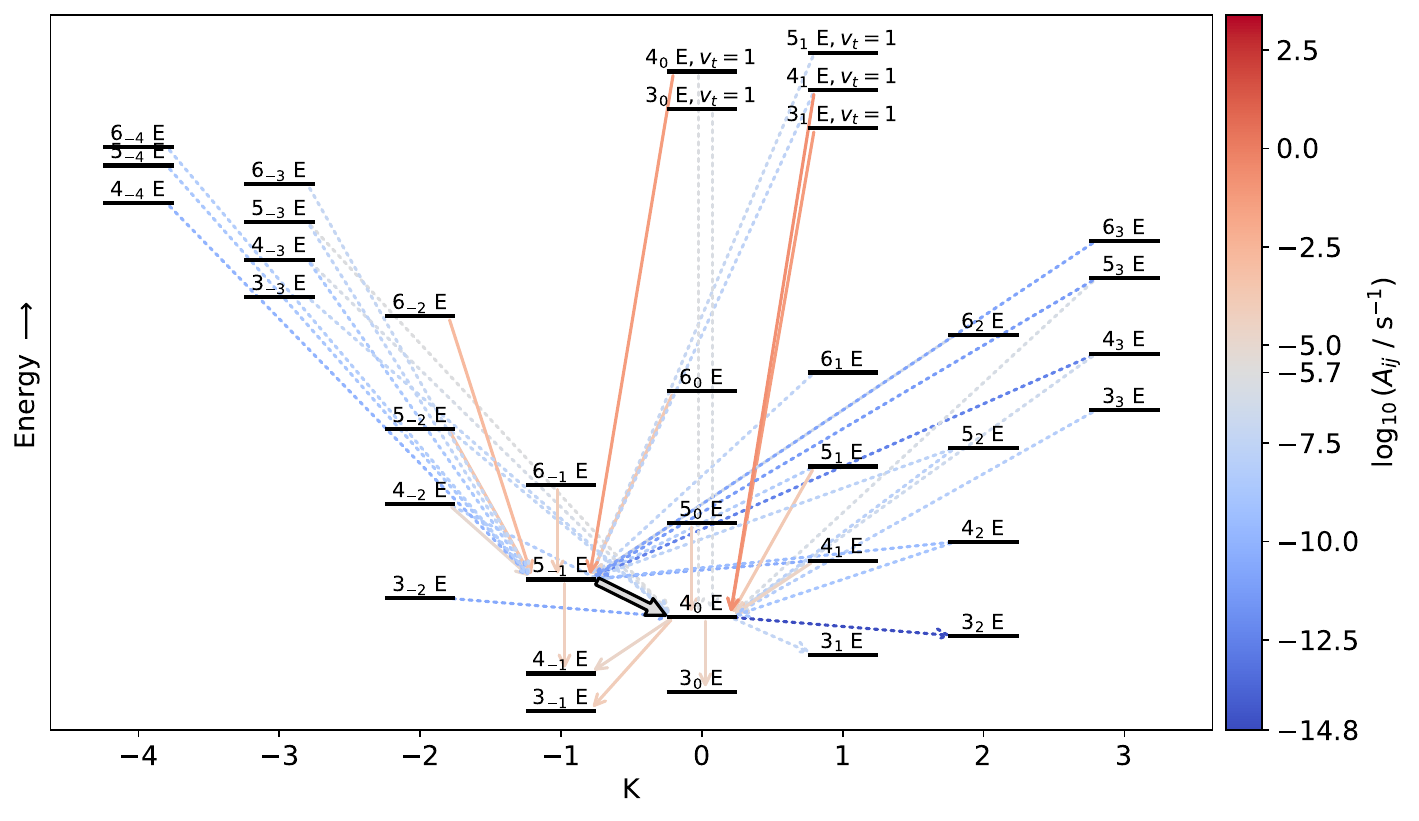}
    \caption{\label{fig:84GHz-levels}
    Energy level structure of the connected rotational transitions for the \ce{CH3OH} $5_{-1} - 4_0\ E$ transition. Energy levels are classified by the $K$ quantum number and ordered by increasing energy but are not drawn to scale. The $\log_{10}{\frac{A_{ij}}{\mathrm{s^{-1}}}}$ for each transition is shown with diverging color schemes, in which the critical middle value is set to be the $\log_{10}{\frac{A_{ij}}{\mathrm{s^{-1}}}}$ of the $5_{-1} - 4_0\ E$ transition~($-5.7$). The targeted transitions are shown with  gray arrows. Transitions with a smaller $A_{ij}$ are shown as blue dotted lines while transitions with a larger $A_{ij}$ are shown as red solid lines.
    }
\end{figure*}

\begin{figure}
    \centering
    \includegraphics[width=0.5\textwidth]{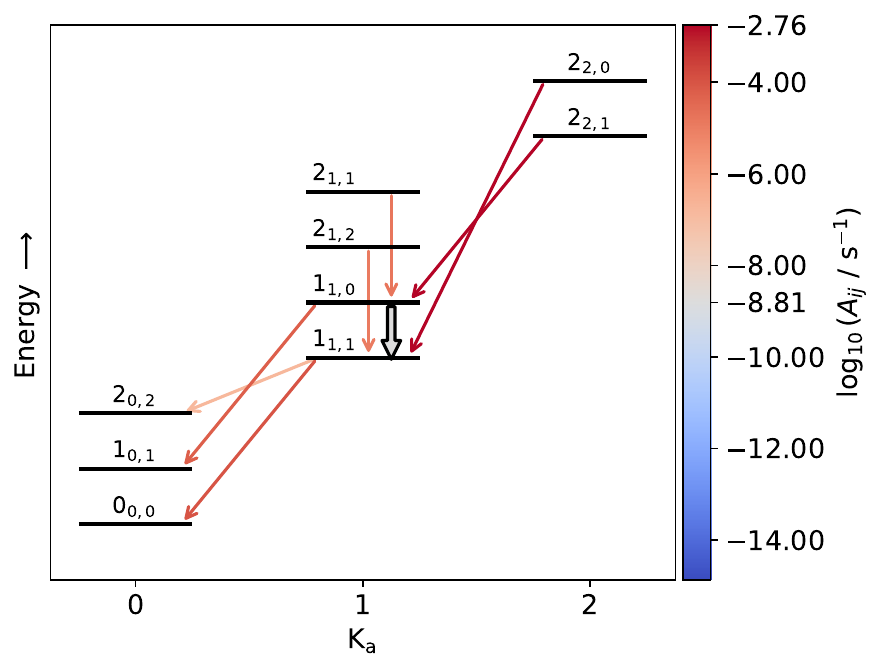}
    \caption{\label{fig:5.3GHz-levels}
    Energy level structure of the connected rotational transitions for \ce{CH2NH} $1_{1, 0} - 1_{1, 1}$ transition at 5288~MHz (thick gray arrow). Similar to Figure~\ref{fig:84GHz-levels}, but the critical middle value is set to be the $\log_{10}{\frac{A_{ij}}{\mathrm{s^{-1}}}}$ of the $1_{1, 0} - 1_{1, 1}$ transition ($-8.81$).
    }
\end{figure}

The majority of the molecular lines observed toward the \ion{H}{2} regions in Sgr B2(N) are in absorption \citep[][and reference therein]{2015MNRAS.452.3969C}. In contrast, specific transitions of certain species toward these regions exhibit emission features \citep[e.g.][]{2012ApJ...758L..33M, 2015MNRAS.452.3969C, 2022A&A...658A..85R}. These transitions establish population inversions between energy levels ($n_{up}g_{low} > n_{low}g_{up}$), a prerequisite for maser action to occur. The mechanism of population inversion is conceptualized by spontaneous decay and collisions: spontaneous decay populates the levels determined by the Einstein coefficients, meanwhile collisions redistribute molecules due to their specific geometry and rotational energy level patterns. Figures~\ref{fig:84GHz-levels} and \ref{fig:5.3GHz-levels} show the energy-level structure of spontaneous transitions connected to the two target maser transitions, with Einstein spontaneous emission coefficients ($A_{ij}$) shown in colors. 

In the case of the \ce{CH3OH} $5_{-1}-4_{0}\ E$ transition at 84~GHz, the upper-energy level $5_{-1}\ E$ is fed by the other higher levels with strong landing transitions but spontaneously decays to the lower-energy level $4_{0}\ E$ with a weak escaping transition \footnote{\footnotesize{An escaping transition refers to a transition in which the upper-energy level is the target energy level, while a landing transition refers to a transition in which the lower-energy level is the target energy level.}}. $5_{-1}\ E$ has a total escaping $\mathrm{log}_{10}A_{ij}/\rm{s}^{-1}$ of $-4.2139$ (equivalent to a lifetime of $\sim1.6 \times 10^4\,$s), which is $\sim$3.0 orders of magnitude lower than its total landing $A_{ij}$. As discussed in \citet{1973ApJ...184..763L} and \citet{1991ASPC...16..119M}, \ce{CH3OH} spontaneously decays more rapidly within the $K=-1$ stack as compared with decaying out of the $K=-1$ stack. Specifically, among the two escaping transitions, $5_{-1}-4_{-1}\ E$ is stronger than $5_{-1}-4_{0}\ E$ (Figures~\ref{fig:84GHz-levels}). In fact, the $K=-1$ levels are the end-points for all spontaneous decay routes \citep{1973ApJ...184..763L}. For example, the lower-energy level of this maser transition ($4_{0}\ E$) decays rapidly to the lower $K=-1$ levels.

In addition to the spontaneous decay, collisions redistribute molecules over the vertical stacks of the $K=-1$ levels because the $\Delta K=0$ collisions are preferred compared with $|\Delta K|=1$ and 2 \citep{1991ASPC...16..119M}. In particular, at a $T_k$ of 30 K, the total rate coefficient of the $\Delta K=0$ collisions de-populating a $K=-1$ level ($\sim1\times10^{-9}\ \mathrm{cm}^3\mathrm{s}^{-1}$) is almost two orders of magnitudes higher than that of the $|\Delta K|=1$ collisions ($\sim5\times10^{-11}\ \mathrm{cm}^3\mathrm{s}^{-1}$) and the $|\Delta K|=2$ collisions ($\sim3\times10^{-11}\ \mathrm{cm}^3\mathrm{s}^{-1}$).

For \ce{CH2NH}, we sum the hyperfine components of the $^{14}$N splittings and use the sum of $A_{ij}$ as the representative $A_{ij}$ for each transition in Figure~\ref{fig:5.3GHz-levels}. In the case of the \ce{CH2NH} $1_{1, 0} - 1_{1, 1}$ transition at 5.29~GHz, the upper-energy level $1_{1, 0}$ is coupled to the higher levels $2_{1, 1}$ and $2_{2, 1}$ with strong spontaneous transitions but to the lower-energy level $1_{1, 1}$ with a weak escaping transition. Compared with the upper-energy level $1_{1, 0}$, which has a total escaping $\mathrm{log}_{10}A_{ij}/\rm{s}^{-1}$ of $-4.2476$, (equivalent to a lifetime of $\sim1.8~\times~10^4\,$s), the lower-energy level $1_{1, 1}$ has a slightly stronger escaping $\mathrm{log}_{10}A_{ij}/\rm{s}^{-1}$ of $-4.0303$ (equivalent to a shorter lifetime of $\sim1.1~\times~10^4\,$s).

Meanwhile, collisions of \ce{CH2NH} exhibit a propensity for the $|\Delta K_a|=0$ transitions compared with the inter-stack transitions $|\Delta K_a|=1$ and 2 \citep{2018JPCL....9.3199F}. In particular, the total rate coefficient of the $\Delta K_a=0$ collisions de-populating a $K_a=1$ level ($\sim10^{-10}\ \mathrm{cm}^3\mathrm{s}^{-1}$) is one order of magnitude higher than that of the $|\Delta K_a|=1$ and 2 collisions at a $T_k$ of 30 K. However, the inter-stack transitions from the ground state $0_{0, 0}$ complicate this multi-level system. The collisions from $0_{0, 0}$ more efficiently populate the lower level $1_{1, 1}$ than the upper level $1_{1, 0}$, suppressing the inversion. Therefore, while these energy structures allow the temporary trapping of the population in the metastable $1_{1, 0}$ level, it only results in a weak population inversion.

\subsection{Spatial Origin and Pumping Mechanism \label{sec:pumpingmechanism}}
\begin{figure}
    \centering
    \includegraphics[width=0.5\textwidth]{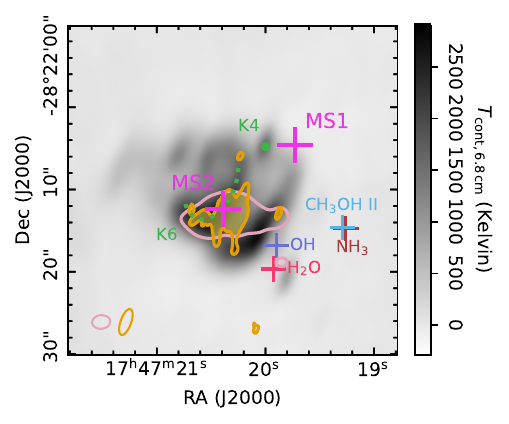}
    \caption{\label{fig:masers-K6}
        The 83-$\mathrm{km\,s^{-1}}$ components of the 84 GHz Class~I \ce{CH3OH} maser and 5.29 GHz \ce{CH2NH} maser are shown with contours in pink and orange respectively, and superimposed on the free-free continuum emission at 6.8 cm. Yellow shading indicates the overlap region where the \ce{CH3OH} emission exceeds $40\,\sigma$ and the \ce{CH2NH} emission exceed $4\,\sigma$. The 84 GHz Class~I \ce{CH3OH} maser spots, MS1 and MS2, are marked with $+$ symbols in megenta while the brightest maser spots of \ce{OH}, \ce{H2O}, \ce{NH3}, and Class~II \ce{CH3OH} masers are marked in other colors \citep{2000ApJS..129..159A, 2010MNRAS.404.1029C, 2014MNRAS.442.2240W, 2022A&A...666L..15Y}. The K4 core and arc-like K6 \ion{H}{2} region are marked with a green circle maker and dotted line respectively.
    }
\end{figure}

The maser activity of several simple species was reported previously toward Sgr B2(N) with precise locations, including \ce{CH3OH}, \ce{H2O}, \ce{OH}, and \ce{NH3}, which are indicated by $+$ symbols in Figures~\ref{fig:maser-obs-sgrb2} and \ref{fig:masers-K6}. The Class~II \ce{CH3OH} maser spot at $6.7\,\mathrm{GHz}$ is spatially associated with the N3 mm-wave core \citep{1996MNRAS.283..606C, 2010MNRAS.404.1029C}. A survey performed with the Australia Telescope Compact Array (ATCA) characterized 30 \ce{H2O} maser spots with the majority of them located within the N1-K2-K3 region \citep{2014MNRAS.442.2240W}. Six \ce{OH} maser spots have been identified with a VLA survey of \ce{OH} masers with the brightest spot being associated with the K3 cm-wave core \citep{2000ApJS..129..159A, 2014MNRAS.441.3137Q}. In a recent VLA observation, nine \ce{NH3} (9,6) maser spots were resolved in this region; most of these spots are concentrated in the N1-K2-K3 region, while the brightest spot is located toward the N3 hot molecular core \citep{2022A&A...666L..15Y}. In contrast, the maser emission of the organic molecular species presented in this work is found to be located close to the K4 cm-wave core and the K6 \ion{H}{2} region, being spatially offset from the spots of the radiative pumping maser species \ce{OH} and Class~II \ce{CH3OH}. The $84\,\mathrm{GHz}$ \ce{CH3OH} maser displays a bright but extended distribution, centered around MS1 (near the K4 core), and subtending an angular size of $\sim15\arcsec$. The emission tapers off rapidly from the center with non-Gaussian profiles, a morphology possibly attributed to maser beaming effects.

We found that MS2, superimposed on the K6 \ion{H}{2} region, harbors maser emission of both Class~I \ce{CH3OH} at $84\,\mathrm{GHz}$ and \ce{CH2NH} at $5.29\,\mathrm{GHz}$ (Figure~\ref{fig:masers-K6}). The 83-$\mathrm{km\,s^{-1}}$ component of the \ce{CH2NH} maser at K6 remarkably resembles that of the Class~I \ce{CH3OH} maser. Their consistent morphology serves as strong evidence to support a similar collisional pumping mechanism for the \ce{CH2NH} maser with the Class~I \ce{CH3OH} maser as well as a common physical condition for triggering these masers. By understanding the conditions that are necessary for maser activities to occur, we can also better understand how they can be used to monitor astronomical objects. For example, since these masers are pumped by intense collisions, we can expect them to be good candidates for studying regions with time-dependent gas dynamics, such as episodic outflows from massive protostars. Future surveys of these new masers in addition to such objects are needed to determine how frequently they accompany Class~I \ce{CH3OH} masers and how they might fit into proposed evolutionary timelines of masers in massive star formation \citep[e.g.,][]{Ellingsen2012}.

Alongside our observations of the Class~I 84 GHz \ce{CH3OH} maser, we also examined the \ce{CH3OH} $8_0-7_1\ A$-type transition at $95\,\mathrm{GHz}$, which exhibits maser activity toward several famous star-forming regions, e.g., Orion–KL \citep{1988ApJ...330L..61P}, DR 21 \citep{1990ApJ...364..555P}, and W51 \citep{1993LNP...412..211P}. A recent single-dish survey of ATLASGAL clumps shows that the $84\,\mathrm{GHz}$ line is generally weaker than the 95\,GHz line \citep{Yang2023}. In contrast, we found that the 95 GHz line exhibits weaker emission toward the identified maser spots in Sgr B2(N) compared to the 84 GHz line. The spatial distribution for the 95 GHz emission is similar to that for other thermal molecular lines, concentrating toward the N1 mm-wave source. This distribution may be attributed to either a weak maser or thermal radiation at different emission spots. It is noteworthy that \citet{1997ApJ...474..346M} identified both maser and quasi-thermal emission components toward Sgr B2 in the 44 GHz line, a consecutive transition ($J=6$) of the $(J+1)_0 - J_1\ A$ series with the 95 GHz transition ($J=7$). A more detailed analysis of other $A$-type \ce{CH3OH} transitions is needed to determine the nature of the 95 GHz emission.

\subsection{Physical Conditions of Masing Clumps \label{sec:physicalconditions}}
The high spatial resolutions of interferometric observations enable us to resolve and segregate compact molecular emissions in diverse environments. In particular, the rich thermalized emission of \ce{CH3OH} and \ce{CH2NH} occurs in the hot core regions. In the N1 hot core, the ALMA observations reported an $N_{col}$ of $2.0\times10^{19}\,\mathrm{cm^{-2}}$ for \ce{CH3OH} and $9.0\times10^{17}\,\mathrm{cm^{-2}}$ for \ce{CH2NH} at a $T_{ex}$ of 230 K \citep{2020A&A...642A..29M, 2022A&A...663A.132M}. In contrast, the bright stimulated maser emission of these two molecules concentrates in the less dense masing clumps. The $N_{col}$ of these two molecules in the masing clumps (Table~\ref{tab:MeOH-58kms-par}-\ref{tab:83kms-par}) are $2-3$ orders of magnitudes less abundant than those in the hot core. 

From the MCMC analyses on the \ce{CH3OH} 84 GHz maser presented in Section~\ref{sec:ch3oh}, we found the $T_k$ for the masing clumps to be ${\sim}29-37\,\mathrm{K}$. This range of $T_k$ is consistent with the large-scale measurement of the Sgr B2 envelope from the OCS lines \citep[30-37 K,][]{2020MNRAS.499.4918A} as well as the dust temperature of Sgr B2(N) obtained from \textit{Herschel} observations \citep[$\sim28$ K,][]{2013A&A...556A.137E}. It is also consistent with the value of $T_k$ (30 K) derived from the non-LTE modeling of the weak masers of methyl formate detected in PRIMOS \citep{2014ApJ...783...72F}.

We also constrain the $n_{\ce{H2}}$ of the masing clumps to be $(5-8)\times10^4\,\mathrm{cm^{-3}}$, a level significantly lower than that derived for the hot core regions. The $n_{\ce{H2}}$ of the Sgr B2(N) hot cores are $>10^7\,\mathrm{cm^{-3}}$, with the N1 core reaching $\sim(1-2)\times10^9\,\mathrm{cm^{-3}}$ \citep{2017A&A...604A...6S}. In contrast, the $n_{\ce{H2}}$ of the masing clumps is comparable to the $n_{\ce{H2}}$ of the colder envelope of Sgr B2, which is estimated to be $10^3-10^4\,\mathrm{cm^{-3}}$ based on the absorption molecular lines \citep{1995A&A...294..667H, 2016A&A...588A.143S}. The slight enhancement in physical densities might be attributed to local energetic events occurring in the envelope necessary for a collisionally pumped maser.

This work presents a quantitative constraint on the physical conditions characterizing a masing clump of organic molecules. It serves as a reference for the physical conditions sufficient for exciting population inversion and collisionally-pumped masers. These conditions include a $T_k$ of $\sim30$, $n_{\ce{H2}}\,>\,5\times10^4\,\mathrm{cm^{-3}}$, and moderate molecular $N_\mathrm{col}$ ($>10^{16}\,\mathrm{cm^{-2}}$ for \ce{CH3OH} and $>10^{14}\,\mathrm{cm^{-2}}$ for \ce{CH2NH}). The relatively low values of temperatures and densities indicate that the masing clumps are situated in the colder and low-density external envelope. The distance from Sgr B2(N) to the edge of the external envelope is suggested to be approximately 19 pc \citep{2016A&A...588A.143S}. The continuum emission from Sgr B2(N) has an angular size of $\sim10\arcsec$, or 0.4 pc at a distance of 8.34 kpc \citep{2014ApJ...783..130R}, which subtends a solid angle of $3.5\times10^{-4}$ sr to the envelope. This aligns with our assumption mentioned in Section~\ref{sec:non-lte}, where background continuum emission sources are far from masing clumps and, therefore, do not significantly affect the local energy distribution of the molecules. Maser activities can manifest themselves in the presence of either a bright continuum background radiation or multiple chained masing clumps in the direction of a diffuse and weak background.

\subsection{Caveat on Geometric Effects \label{sec:caveat}}
Geometric effects related to the propagation of maser radiation become increasingly significant as the optical depth becomes lower than -1. Although \texttt{RADEX} and \texttt{MOLSIM} support various geometries through escape probability approximation, other effects such as maser beaming are not treated. Such effects are particularly important in the case of unresolved clumps. The models also assume homogeneous physical conditions (i.e. one-zone models), which is inappropriate for very negative optical depths. For precisely modeling maser lines with strong amplification, a non-local treatment of radiation and level population is required.

Although our current treatment does not provide a precise solution, it is in principle appropriate for $\mid\tau\mid\ll1$ (the case of \ce{CH2NH}) and still offers a reasonable approximation to the first order as $\mid\tau\mid$ is of the order of unity (the case of \ce{CH3OH}). Incorporating corrections such as non-local treatments and other geometric effects would definitely improve the solution but it would also introduce significant computational costs. This becomes especially infeasible for MCMC calculations. The current solution represents the best that is achievable with our current resources, acknowledging that this approach is subject to future expansions and refinements.

\section{Summary}
We report interferometric observations of maser activity from organic molecules toward Sgr B2 using the ALMA and VLA. Enabled by the ALMA observations, we report the first detection of the Class~I \ce{CH3OH} maser at $84\,\mathrm{GHz}$ toward Sgr B2. Multiple $84\,\mathrm{GHz}$~\ce{CH3OH} maser knots are discovered with the brightest spot MS1 located ${\sim}3\arcsec$ west of the K4 free-free continuum emission core. The peak brightness temperature of MS1 reaches $389.2\pm0.2\,\mathrm{K}$ at a velocity of $58\ \mathrm{km\,s^{-1}}$. In addition, another \ce{CH3OH} maser spot MS2 coexists with the K6 UC \ion{H}{2} region at a velocity of $83\ \mathrm{km\,s^{-1}}$. With the VLA observations, we resolve the $5.29\,\mathrm{GHz}$ \ce{CH2NH} maser emission into three velocity components. These three components display distinct fine-scale structures associated with the K5 and K6 bow-shaped \ion{H}{2} regions. The $83\ \mathrm{km\,s^{-1}}$ component is co-spatial with MS2 and its peak brightness temperature has a lower limit of $121.8\pm8.6\,\mathrm{K}$. 

We present a tool for modeling and fitting maser spectra with non-LTE calculations and a Bayesian approach. By leveraging the theoretical calculations on collisional rate coefficients and cross sections for both $E$-type \ce{CH3OH} and \ce{CH2NH} transitions, this fitting technique enables us to statistically constrain the physical conditions of the masing regions. In particular, our analysis suggests a two-chain-clump maser model for explaining the intense \ce{CH3OH} maser activity observed toward a region with weak and diffuse background radiation. This model can reproduce both the maser and the optically thick emission features of the \ce{CH3OH} spectra observed toward the brightest maser spot MS1.

Interferometric observations have not only enabled us to characterize the spatial origin and extent of maser emission but also helped to constrain the pumping mechanism through direct comparisons of their morphologies. We found a definite association between the maser emission from complex molecules and the K6 UC \ion{H}{2} region. They are offset from the masing regions of the radiatively pumped maser species and the infrared sources. In contrast, the spatial correlation between the activities of the 84 GHz \ce{CH3OH} Class~I maser and 5.29 GHz \ce{CH2NH} maser suggest that the \ce{CH2NH} maser is pumped by intense collisions, analogous to the Class~I \ce{CH3OH} maser. By understanding the nature of the new maser species, which are pumped by intense collisions, we can better understand how they may serve as good candidates for studying and monitoring astronomical objects with time-dependent gas dynamics.

\section{Data access \& code}
The code used to perform the analysis is part of the \texttt{molsim} open-source package; an archival version of the code can be accessed at \citet{molsim}. The spectroscopic and collisional data for $E$-\ce{CH3OH} and \ce{CH2NH} can be found in the EMAA database (https://emaa.osug.fr and https://dx.doi.org/10.17178/EMAA).

\facilities{ALMA, VLA, GBT}

\software{
    \texttt{Molsim} \citep{molsim},
    \texttt{Emcee} \citep{foreman-mackey_emcee_2013},
    \texttt{Numba} \citep{2015llvm.confE...1L},
    \texttt{CASA} \citep{CASA2022},
          }

\acknowledgments
B. A. M. and C. X. gratefully acknowledge support of National Science Foundation grant AST-2205126. This work was completed in part while C.X. was a Grote Reber Fellow. C.X. acknowledges support from the National Science Foundation through the Grote Reber Fellowship Program administered by Associated Universities, Inc./National Radio Astronomy Observatory and the Virginia Space Grant Consortium. The National Radio Astronomy Observatory is a facility of the National Science Foundation operated under cooperative agreement by Associated Universities, Inc. ALMA is a partnership of ESO (representing its member states), NSF (USA) and NINS (Japan), together with NRC (Canada), MOST and ASIAA (Taiwan), and KASI (Republic of Korea), in cooperation with the Republic of Chile. The Joint ALMA Observatory is operated by ESO, AUI/NRAO and NAOJ.   This paper makes use of the following ALMA data: ADS/JAO.ALMA\#2011.0.00017.S. This research has made use of spectroscopic and collisional data from the EMAA database (https://emaa.osug.fr and https://dx.doi.org/10.17178/EMAA). EMAA is supported by the Observatoire des Sciences de l’Univers de Grenoble (OSUG).

\bibliographystyle{aasjournal}

\begin{thebibliography}{}
\expandafter\ifx\csname natexlab\endcsname\relax\def\natexlab#1{#1}\fi
\providecommand{\url}[1]{\href{#1}{#1}}
\providecommand{\dodoi}[1]{doi:~\href{http://doi.org/#1}{\nolinkurl{#1}}}
\providecommand{\doeprint}[1]{\href{http://ascl.net/#1}{\nolinkurl{http://ascl.net/#1}}}
\providecommand{\doarXiv}[1]{\href{https://arxiv.org/abs/#1}{\nolinkurl{https://arxiv.org/abs/#1}}}

\bibitem[{{Argon} {et~al.}(2000){Argon}, {Reid}, \& {Menten}}]{2000ApJS..129..159A}
{Argon}, A.~L., {Reid}, M.~J., \& {Menten}, K.~M. 2000, \apjs, 129, 159, \dodoi{10.1086/313406}

\bibitem[{{Armijos-Abenda{\~n}o} {et~al.}(2020){Armijos-Abenda{\~n}o}, {Banda-Barrag{\'a}n}, {Mart{\'\i}n-Pintado}, {D{\'e}nes}, {Federrath}, \& {Requena-Torres}}]{2020MNRAS.499.4918A}
{Armijos-Abenda{\~n}o}, J., {Banda-Barrag{\'a}n}, W.~E., {Mart{\'\i}n-Pintado}, J., {et~al.} 2020, \mnras, 499, 4918, \dodoi{10.1093/mnras/staa3119}

\bibitem[{{Batrla} {et~al.}(1987){Batrla}, {Matthews}, {Menten}, \& {Walmsley}}]{1987Natur.326...49B}
{Batrla}, W., {Matthews}, H.~E., {Menten}, K.~M., \& {Walmsley}, C.~M. 1987, \nat, 326, 49, \dodoi{10.1038/326049a0}

\bibitem[{{Baudry} {et~al.}(1974){Baudry}, {Forster}, \& {Welch}}]{1974A&A....36..217B}
{Baudry}, A., {Forster}, J.~R., \& {Welch}, W.~J. 1974, \aap, 36, 217

\bibitem[{{Belloche} {et~al.}(2016){Belloche}, {M{\"u}ller}, {Garrod}, \& {Menten}}]{2016A&A...587A..91B}
{Belloche}, A., {M{\"u}ller}, H.~S.~P., {Garrod}, R.~T., \& {Menten}, K.~M. 2016, \aap, 587, A91, \dodoi{10.1051/0004-6361/201527268}

\bibitem[{{Bonfand} {et~al.}(2017){Bonfand}, {Belloche}, {Menten}, {Garrod}, \& {M{\"u}ller}}]{2017A&A...604A..60B}
{Bonfand}, M., {Belloche}, A., {Menten}, K.~M., {Garrod}, R.~T., \& {M{\"u}ller}, H.~S.~P. 2017, \aap, 604, A60, \dodoi{10.1051/0004-6361/201730648}

\bibitem[{{Breen} {et~al.}(2019){Breen}, {Contreras}, {Dawson}, {Ellingsen}, {Voronkov}, \& {McCarthy}}]{2019MNRAS.484.5072B}
{Breen}, S.~L., {Contreras}, Y., {Dawson}, J.~R., {et~al.} 2019, \mnras, 484, 5072, \dodoi{10.1093/mnras/stz192}

\bibitem[{{Breen} {et~al.}(2010){Breen}, {Ellingsen}, {Caswell}, \& {Lewis}}]{Breen2010}
{Breen}, S.~L., {Ellingsen}, S.~P., {Caswell}, J.~L., \& {Lewis}, B.~E. 2010, \mnras, 401, 2219, \dodoi{10.1111/j.1365-2966.2009.15831.x}

\bibitem[{{Breen} {et~al.}(2013){Breen}, {Ellingsen}, {Contreras}, {Green}, {Caswell}, {Stevens}, {Dawson}, \& {Voronkov}}]{Breen2013}
{Breen}, S.~L., {Ellingsen}, S.~P., {Contreras}, Y., {et~al.} 2013, \mnras, 435, 524, \dodoi{10.1093/mnras/stt1315}

\bibitem[{{CASA Team} {et~al.}(2022){CASA Team}, {Bean}, {Bhatnagar}, {Castro}, {Donovan Meyer}, {Emonts}, {Garcia}, {Garwood}, {Golap}, {Gonzalez Villalba}, {Harris}, {Hayashi}, {Hoskins}, {Hsieh}, {Jagannathan}, {Kawasaki}, {Keimpema}, {Kettenis}, {Lopez}, {Marvil}, {Masters}, {McNichols}, {Mehringer}, {Miel}, {Moellenbrock}, {Montesino}, {Nakazato}, {Ott}, {Petry}, {Pokorny}, {Raba}, {Rau}, {Schiebel}, {Schweighart}, {Sekhar}, {Shimada}, {Small}, {Steeb}, {Sugimoto}, {Suoranta}, {Tsutsumi}, {van Bemmel}, {Verkouter}, {Wells}, {Xiong}, {Szomoru}, {Griffith}, {Glendenning}, \& {Kern}}]{CASA2022}
{CASA Team}, {Bean}, B., {Bhatnagar}, S., {et~al.} 2022, \pasp, 134, 114501, \dodoi{10.1088/1538-3873/ac9642}

\bibitem[{{Caswell}(1996)}]{1996MNRAS.283..606C}
{Caswell}, J.~L. 1996, \mnras, 283, 606, \dodoi{10.1093/mnras/283.2.606}

\bibitem[{{Caswell} {et~al.}(2010){Caswell}, {Fuller}, {Green}, {Avison}, {Breen}, {Brooks}, {Burton}, {Chrysostomou}, {Cox}, {Diamond}, {Ellingsen}, {Gray}, {Hoare}, {Masheder}, {McClure-Griffiths}, {Pestalozzi}, {Phillips}, {Quinn}, {Thompson}, {Voronkov}, {Walsh}, {Ward-Thompson}, {Wong-McSweeney}, {Yates}, \& {Cohen}}]{2010MNRAS.404.1029C}
{Caswell}, J.~L., {Fuller}, G.~A., {Green}, J.~A., {et~al.} 2010, \mnras, 404, 1029, \dodoi{10.1111/j.1365-2966.2010.16339.x}

\bibitem[{{Chen} {et~al.}(2020){Chen}, {Sobolev}, {Ren}, {Parfenov}, {Breen}, {Ellingsen}, {Shen}, {Li}, {MacLeod}, {Baan}, {Brogan}, {Hirota}, {Hunter}, {Linz}, {Menten}, {Sugiyama}, {Stecklum}, {Gong}, \& {Zheng}}]{2020NatAs...4.1170C}
{Chen}, X., {Sobolev}, A.~M., {Ren}, Z.-Y., {et~al.} 2020, Nature Astronomy, 4, 1170, \dodoi{10.1038/s41550-020-1144-x}

\bibitem[{{Corby} {et~al.}(2015){Corby}, {Jones}, {Cunningham}, {Menten}, {Belloche}, {Schwab}, {Walsh}, {Balnozan}, {Bronfman}, {Lo}, \& {Remijan}}]{2015MNRAS.452.3969C}
{Corby}, J.~F., {Jones}, P.~A., {Cunningham}, M.~R., {et~al.} 2015, \mnras, 452, 3969, \dodoi{10.1093/mnras/stv1494}

\bibitem[{{Cragg} {et~al.}(2002){Cragg}, {Sobolev}, \& {Godfrey}}]{2002MNRAS.331..521C}
{Cragg}, D.~M., {Sobolev}, A.~M., \& {Godfrey}, P.~D. 2002, \mnras, 331, 521, \dodoi{10.1046/j.1365-8711.2002.05226.x}

\bibitem[{{Cragg} {et~al.}(2005){Cragg}, {Sobolev}, \& {Godfrey}}]{2005MNRAS.360..533C}
---. 2005, \mnras, 360, 533, \dodoi{10.1111/j.1365-2966.2005.09077.x}

\bibitem[{{Cyganowski} {et~al.}(2009){Cyganowski}, {Brogan}, {Hunter}, \& {Churchwell}}]{2009ApJ...702.1615C}
{Cyganowski}, C.~J., {Brogan}, C.~L., {Hunter}, T.~R., \& {Churchwell}, E. 2009, \apj, 702, 1615, \dodoi{10.1088/0004-637X/702/2/1615}

\bibitem[{{Cyganowski} {et~al.}(2012){Cyganowski}, {Brogan}, {Hunter}, {Zhang}, {Friesen}, {Indebetouw}, \& {Chandler}}]{2012ApJ...760L..20C}
{Cyganowski}, C.~J., {Brogan}, C.~L., {Hunter}, T.~R., {et~al.} 2012, \apjl, 760, L20, \dodoi{10.1088/2041-8205/760/2/L20}

\bibitem[{{Elitzur} \& {Fuqua}(1989)}]{1989ApJ...347L..35E}
{Elitzur}, M., \& {Fuqua}, J.~B. 1989, \apjl, 347, L35, \dodoi{10.1086/185601}

\bibitem[{{Ellingsen}(2006)}]{2006ApJ...638..241E}
{Ellingsen}, S.~P. 2006, \apj, 638, 241, \dodoi{10.1086/498673}

\bibitem[{{Ellingsen} {et~al.}(2012){Ellingsen}, {Breen}, {Voronkov}, {Caswell}, {Chen}, \& {Titmarsh}}]{Ellingsen2012}
{Ellingsen}, S.~P., {Breen}, S.~L., {Voronkov}, M.~A., {et~al.} 2012, arXiv e-prints, arXiv:1210.2139, \dodoi{10.48550/arXiv.1210.2139}

\bibitem[{{Etxaluze} {et~al.}(2013){Etxaluze}, {Goicoechea}, {Cernicharo}, {Polehampton}, {Noriega-Crespo}, {Molinari}, {Swinyard}, {Wu}, \& {Bally}}]{2013A&A...556A.137E}
{Etxaluze}, M., {Goicoechea}, J.~R., {Cernicharo}, J., {et~al.} 2013, \aap, 556, A137, \dodoi{10.1051/0004-6361/201321258}

\bibitem[{{Faure} {et~al.}(2018){Faure}, {Lique}, \& {Remijan}}]{2018JPCL....9.3199F}
{Faure}, A., {Lique}, F., \& {Remijan}, A.~J. 2018, The Journal of Physical Chemistry Letters, 9, 3199, \dodoi{10.1021/acs.jpclett.8b01431}

\bibitem[{{Faure} {et~al.}(2014){Faure}, {Remijan}, {Szalewicz}, \& {Wiesenfeld}}]{2014ApJ...783...72F}
{Faure}, A., {Remijan}, A.~J., {Szalewicz}, K., \& {Wiesenfeld}, L. 2014, \apj, 783, 72, \dodoi{10.1088/0004-637X/783/2/72}

\bibitem[{{Foreman-Mackey} {et~al.}(2013){Foreman-Mackey}, {Hogg}, {Lang}, \& {Goodman}}]{foreman-mackey_emcee_2013}
{Foreman-Mackey}, D., {Hogg}, D.~W., {Lang}, D., \& {Goodman}, J. 2013, \pasp, 125, 306, \dodoi{10.1086/670067}

\bibitem[{{Forster} \& {Caswell}(1989)}]{1989A&A...213..339F}
{Forster}, J.~R., \& {Caswell}, J.~L. 1989, \aap, 213, 339

\bibitem[{{Fujisawa} {et~al.}(2014){Fujisawa}, {Sugiyama}, {Motogi}, {Hachisuka}, {Yonekura}, {Sawada-Satoh}, {Matsumoto}, {Sorai}, {Momose}, {Saito}, {Takaba}, {Ogawa}, {Kimura}, {Niinuma}, {Hirano}, {Omodaka}, {Kobayashi}, {Kawaguchi}, {Shibata}, {Honma}, {Hirota}, {Murata}, {Doi}, {Mochizuki}, {Shen}, {Chen}, {Xia}, {Li}, \& {Kim}}]{2014PASJ...66...31F}
{Fujisawa}, K., {Sugiyama}, K., {Motogi}, K., {et~al.} 2014, \pasj, 66, 31, \dodoi{10.1093/pasj/psu015}

\bibitem[{{Gargaud} {et~al.}(2011){Gargaud}, {Amils}, {Quintanilla}, {Cleaves}, {Irvine}, {Pinti}, \& {Viso}}]{2011eab..book.....G}
{Gargaud}, M., {Amils}, R., {Quintanilla}, J.~C., {et~al.} 2011, {Encyclopedia of Astrobiology} (Springer Berlin Heidelberg), \dodoi{10.1007/978-3-642-11274-4}

\bibitem[{{Gorski} {et~al.}(2021){Gorski}, {Aalto}, {Mangum}, {Momjian}, {Black}, {Falstad}, {Gullberg}, {K{\"o}nig}, {Onishi}, {Sato}, \& {Stanley}}]{2021A&A...654A.110G}
{Gorski}, M.~D., {Aalto}, S., {Mangum}, J., {et~al.} 2021, \aap, 654, A110, \dodoi{10.1051/0004-6361/202141633}

\bibitem[{{Hoffman} {et~al.}(2007){Hoffman}, {Goss}, \& {Palmer}}]{2007ApJ...654..971H}
{Hoffman}, I.~M., {Goss}, W.~M., \& {Palmer}, P. 2007, \apj, 654, 971, \dodoi{10.1086/509717}

\bibitem[{{Huettemeister} {et~al.}(1995){Huettemeister}, {Wilson}, {Mauersberger}, {Lemme}, {Dahmen}, \& {Henkel}}]{1995A&A...294..667H}
{Huettemeister}, S., {Wilson}, T.~L., {Mauersberger}, R., {et~al.} 1995, \aap, 294, 667

\bibitem[{{Humire} {et~al.}(2022){Humire}, {Henkel}, {Hern{\'a}ndez-G{\'o}mez}, {Mart{\'\i}n}, {Mangum}, {Harada}, {Muller}, {Sakamoto}, {Tanaka}, {Yoshimura}, {Nakanishi}, {M{\"u}hle}, {Herrero-Illana}, {Meier}, {Caux}, {Aladro}, {Mauersberger}, {Viti}, {Colzi}, {Rivilla}, {Gorski}, {Menten}, {Huang}, {Aalto}, {van der Werf}, \& {Emig}}]{Humire2022}
{Humire}, P.~K., {Henkel}, C., {Hern{\'a}ndez-G{\'o}mez}, A., {et~al.} 2022, \aap, 663, A33, \dodoi{10.1051/0004-6361/202243384}

\bibitem[{{Hunter} {et~al.}(2018){Hunter}, {Brogan}, {Bartkiewicz}, {Chibueze}, {Cyganowski}, {Hirota}, {MacLeod}, {Sanna}, \& {Torrelles}}]{2018ASPC..517..321H}
{Hunter}, T.~R., {Brogan}, C.~L., {Bartkiewicz}, A., {et~al.} 2018, in Astronomical Society of the Pacific Conference Series, Vol. 517, Science with a Next Generation Very Large Array, ed. E.~{Murphy}, 321

\bibitem[{{Hutson} \& {Green}(2012)}]{molscat}
{Hutson}, J.~M., \& {Green}, S. 2012, {MOLSCAT: MOLecular SCATtering}, Astrophysics Source Code Library.
\newblock \doeprint{1206.004}

\bibitem[{{Kartje} {et~al.}(1999){Kartje}, {K{\"o}nigl}, \& {Elitzur}}]{1999ApJ...513..180K}
{Kartje}, J.~F., {K{\"o}nigl}, A., \& {Elitzur}, M. 1999, \apj, 513, 180, \dodoi{10.1086/306824}

\bibitem[{{Lam} {et~al.}(2015){Lam}, {Pitrou}, \& {Seibert}}]{2015llvm.confE...1L}
{Lam}, S.~K., {Pitrou}, A., \& {Seibert}, S. 2015, in Proc. Second Workshop on the LLVM Compiler Infrastructure in HPC, 1--6

\bibitem[{Lee {et~al.}(2023)Lee, Loomis, Xue, El-Abd, \& McGuire}]{molsim}
Lee, K. L.~K., Loomis, R.~A., Xue, C., El-Abd, S., \& McGuire, B.~A. 2023, molsim,  Zenodo, \dodoi{10.5281/ZENODO.8118192}.
\newblock \url{https://doi.org/10.5281/zenodo.8118192}

\bibitem[{{Lees}(1973)}]{1973ApJ...184..763L}
{Lees}, R.~M. 1973, \apj, 184, 763, \dodoi{10.1086/152368}

\bibitem[{{Leurini} {et~al.}(2016){Leurini}, {Menten}, \& {Walmsley}}]{2016A&A...592A..31L}
{Leurini}, S., {Menten}, K.~M., \& {Walmsley}, C.~M. 2016, \aap, 592, A31, \dodoi{10.1051/0004-6361/201527974}

\bibitem[{{Margul{\`e}s} {et~al.}(2022){Margul{\`e}s}, {Remijan}, {Belloche}, {Motiyenko}, {McGuire}, {Xue}, {M{\"u}ller}, {Garrod}, {Menten}, \& {Guillemin}}]{2022A&A...663A.132M}
{Margul{\`e}s}, L., {Remijan}, A., {Belloche}, A., {et~al.} 2022, \aap, 663, A132, \dodoi{10.1051/0004-6361/202243172}

\bibitem[{{Mather} {et~al.}(1999){Mather}, {Fixsen}, {Shafer}, {Mosier}, \& {Wilkinson}}]{Mather1999}
{Mather}, J.~C., {Fixsen}, D.~J., {Shafer}, R.~A., {Mosier}, C., \& {Wilkinson}, D.~T. 1999, \apj, 512, 511, \dodoi{10.1086/306805}

\bibitem[{{McCarthy} {et~al.}(2018){McCarthy}, {Ellingsen}, {Breen}, {Voronkov}, \& {Chen}}]{Tiege2018}
{McCarthy}, T.~P., {Ellingsen}, S.~P., {Breen}, S.~L., {Voronkov}, M.~A., \& {Chen}, X. 2018, \apjl, 867, L4, \dodoi{10.3847/2041-8213/aae82c}

\bibitem[{{McGrath} {et~al.}(2004){McGrath}, {Goss}, \& {De Pree}}]{2004ApJS..155..577M}
{McGrath}, E.~J., {Goss}, W.~M., \& {De Pree}, C.~G. 2004, \apjs, 155, 577, \dodoi{10.1086/424486}

\bibitem[{{McGuire} {et~al.}(2012){McGuire}, {Loomis}, {Charness}, {Corby}, {Blake}, {Hollis}, {Lovas}, {Jewell}, \& {Remijan}}]{2012ApJ...758L..33M}
{McGuire}, B.~A., {Loomis}, R.~A., {Charness}, C.~M., {et~al.} 2012, \apjl, 758, L33, \dodoi{10.1088/2041-8205/758/2/L33}

\bibitem[{{Mehringer} \& {Menten}(1997)}]{1997ApJ...474..346M}
{Mehringer}, D.~M., \& {Menten}, K.~M. 1997, \apj, 474, 346, \dodoi{10.1086/303454}

\bibitem[{{Menten}(1991)}]{1991ASPC...16..119M}
{Menten}, K.~M. 1991, in Astronomical Society of the Pacific Conference Series, Vol.~16, Atoms, Ions and Molecules: New Results in Spectral Line Astrophysics, ed. A.~D. {Haschick} \& P.~T.~P. {Ho}, 119--136

\bibitem[{{Morita} {et~al.}(1992){Morita}, {Hasegawa}, {Ukita}, {Okumura}, \& {Ishiguro}}]{1992PASJ...44..373M}
{Morita}, K.-I., {Hasegawa}, T., {Ukita}, N., {Okumura}, S.~K., \& {Ishiguro}, M. 1992, \pasj, 44, 373

\bibitem[{{Moscadelli} {et~al.}(2016){Moscadelli}, {S{\'a}nchez-Monge}, {Goddi}, {Li}, {Sanna}, {Cesaroni}, {Pestalozzi}, {Molinari}, \& {Reid}}]{2016A&A...585A..71M}
{Moscadelli}, L., {S{\'a}nchez-Monge}, {\'A}., {Goddi}, C., {et~al.} 2016, \aap, 585, A71, \dodoi{10.1051/0004-6361/201526238}

\bibitem[{{Motiyenko} {et~al.}(2020){Motiyenko}, {Belloche}, {Garrod}, {Margul{\`e}s}, {M{\"u}ller}, {Menten}, \& {Guillemin}}]{2020A&A...642A..29M}
{Motiyenko}, R.~A., {Belloche}, A., {Garrod}, R.~T., {et~al.} 2020, \aap, 642, A29, \dodoi{10.1051/0004-6361/202038723}

\bibitem[{{M{\"u}ller} {et~al.}(2005){M{\"u}ller}, {Schl{\"o}der}, {Stutzki}, \& {Winnewisser}}]{2005JMoSt.742..215M}
{M{\"u}ller}, H. S.~P., {Schl{\"o}der}, F., {Stutzki}, J., \& {Winnewisser}, G. 2005, Journal of Molecular Structure, 742, 215, \dodoi{10.1016/j.molstruc.2005.01.027}

\bibitem[{{Plambeck} \& {Menten}(1990)}]{1990ApJ...364..555P}
{Plambeck}, R.~L., \& {Menten}, K.~M. 1990, \apj, 364, 555, \dodoi{10.1086/169437}

\bibitem[{{Plambeck} \& {Wright}(1988)}]{1988ApJ...330L..61P}
{Plambeck}, R.~L., \& {Wright}, M.~C.~H. 1988, \apjl, 330, L61, \dodoi{10.1086/185205}

\bibitem[{{Pratap} \& {Menten}(1993)}]{1993LNP...412..211P}
{Pratap}, P., \& {Menten}, K. 1993, {Interferometric observations of 95 GHz methanol masers.}, Vol. 412 (Springer Berlin Heidelberg), 211--214

\bibitem[{{Qiao} {et~al.}(2014){Qiao}, {Li}, {Shen}, {Chen}, \& {Zheng}}]{2014MNRAS.441.3137Q}
{Qiao}, H., {Li}, J., {Shen}, Z., {Chen}, X., \& {Zheng}, X. 2014, \mnras, 441, 3137, \dodoi{10.1093/mnras/stu776}

\bibitem[{{Qiao} {et~al.}(2018){Qiao}, {Walsh}, {Breen}, {G{\'o}mez}, {Dawson}, {Imai}, {Ellingsen}, {Green}, \& {Shen}}]{2018ApJS..239...15Q}
{Qiao}, H.-H., {Walsh}, A.~J., {Breen}, S.~L., {et~al.} 2018, \apjs, 239, 15, \dodoi{10.3847/1538-4365/aae580}

\bibitem[{{Rabli} \& {Flower}(2010)}]{2010MNRAS.406...95R}
{Rabli}, D., \& {Flower}, D.~R. 2010, \mnras, 406, 95, \dodoi{10.1111/j.1365-2966.2010.16671.x}

\bibitem[{{Reid} {et~al.}(2014){Reid}, {Menten}, {Brunthaler}, {Zheng}, {Dame}, {Xu}, {Wu}, {Zhang}, {Sanna}, \& {Sato}}]{2014ApJ...783..130R}
{Reid}, M.~J., {Menten}, K.~M., {Brunthaler}, A., {et~al.} 2014, \apj, 783, 130, \dodoi{10.1088/0004-637X/783/2/130}

\bibitem[{{Reid} {et~al.}(2019){Reid}, {Menten}, {Brunthaler}, {Zheng}, {Dame}, {Xu}, {Li}, {Sakai}, {Wu}, {Immer}, {Zhang}, {Sanna}, {Moscadelli}, {Rygl}, {Bartkiewicz}, {Hu}, {Quiroga-Nu{\~n}ez}, \& {van Langevelde}}]{2019ApJ...885..131R}
---. 2019, \apj, 885, 131, \dodoi{10.3847/1538-4357/ab4a11}

\bibitem[{{Remijan} {et~al.}(2022){Remijan}, {Xue}, {Margul{\`e}s}, {Belloche}, {Motiyenko}, {Carder}, {Codella}, {Balucani}, {Brogan}, {Ceccarelli}, {Hunter}, {Maris}, {Melandri}, {Siebert}, \& {McGuire}}]{2022A&A...658A..85R}
{Remijan}, A., {Xue}, C., {Margul{\`e}s}, L., {et~al.} 2022, \aap, 658, A85, \dodoi{10.1051/0004-6361/202142504}

\bibitem[{{S{\'a}nchez-Monge} {et~al.}(2017){S{\'a}nchez-Monge}, {Schilke}, {Schmiedeke}, {Ginsburg}, {Cesaroni}, {Lis}, {Qin}, {M{\"u}ller}, {Bergin}, {Comito}, \& {M{\"o}ller}}]{2017A&A...604A...6S}
{S{\'a}nchez-Monge}, {\'A}., {Schilke}, P., {Schmiedeke}, A., {et~al.} 2017, \aap, 604, A6, \dodoi{10.1051/0004-6361/201730426}

\bibitem[{{Sanna} {et~al.}(2016){Sanna}, {Moscadelli}, {Cesaroni}, {Caratti o Garatti}, {Goddi}, \& {Carrasco-Gonz{\'a}lez}}]{2016A&A...596L...2S}
{Sanna}, A., {Moscadelli}, L., {Cesaroni}, R., {et~al.} 2016, \aap, 596, L2, \dodoi{10.1051/0004-6361/201629544}

\bibitem[{{Schmiedeke} {et~al.}(2016){Schmiedeke}, {Schilke}, {M{\"o}ller}, {S{\'a}nchez-Monge}, {Bergin}, {Comito}, {Csengeri}, {Lis}, {Molinari}, {Qin}, \& {Rolffs}}]{2016A&A...588A.143S}
{Schmiedeke}, A., {Schilke}, P., {M{\"o}ller}, T., {et~al.} 2016, \aap, 588, A143, \dodoi{10.1051/0004-6361/201527311}

\bibitem[{{Sobolev} {et~al.}(1997){Sobolev}, {Cragg}, \& {Godfrey}}]{1997A&A...324..211S}
{Sobolev}, A.~M., {Cragg}, D.~M., \& {Godfrey}, P.~D. 1997, \aap, 324, 211

\bibitem[{{Sobolev} \& {Deguchi}(1994)}]{1994A&A...291..569S}
{Sobolev}, A.~M., \& {Deguchi}, S. 1994, \aap, 291, 569

\bibitem[{{Song} {et~al.}(2022){Song}, {Chen}, {Shen}, {Li}, {Yang}, {Ouyang}, {Sobolev}, {Zhao}, {Li}, \& {Cai}}]{2022ApJS..258...19S}
{Song}, S.-M., {Chen}, X., {Shen}, Z.-Q., {et~al.} 2022, \apjs, 258, 19, \dodoi{10.3847/1538-4365/ac348e}

\bibitem[{{Szymczak} {et~al.}(2005){Szymczak}, {Pillai}, \& {Menten}}]{2005A&A...434..613S}
{Szymczak}, M., {Pillai}, T., \& {Menten}, K.~M. 2005, \aap, 434, 613, \dodoi{10.1051/0004-6361:20042437}

\bibitem[{{Towner} {et~al.}(2017){Towner}, {Brogan}, {Hunter}, {Cyganowski}, {McGuire}, {Indebetouw}, {Friesen}, \& {Chandler}}]{2017ApJS..230...22T}
{Towner}, A.~P.~M., {Brogan}, C.~L., {Hunter}, T.~R., {et~al.} 2017, \apjs, 230, 22, \dodoi{10.3847/1538-4365/aa73d8}

\bibitem[{{van der Tak} {et~al.}(2007){van der Tak}, {Black}, {Sch{\"o}ier}, {Jansen}, \& {van Dishoeck}}]{2007A&A...468..627V}
{van der Tak}, F.~F.~S., {Black}, J.~H., {Sch{\"o}ier}, F.~L., {Jansen}, D.~J., \& {van Dishoeck}, E.~F. 2007, \aap, 468, 627, \dodoi{10.1051/0004-6361:20066820}

\bibitem[{{Voronkov} {et~al.}(2006){Voronkov}, {Brooks}, {Sobolev}, {Ellingsen}, {Ostrovskii}, \& {Caswell}}]{2006MNRAS.373..411V}
{Voronkov}, M.~A., {Brooks}, K.~J., {Sobolev}, A.~M., {et~al.} 2006, \mnras, 373, 411, \dodoi{10.1111/j.1365-2966.2006.11047.x}

\bibitem[{{Voronkov} {et~al.}(2014){Voronkov}, {Caswell}, {Ellingsen}, {Green}, \& {Breen}}]{2014MNRAS.439.2584V}
{Voronkov}, M.~A., {Caswell}, J.~L., {Ellingsen}, S.~P., {Green}, J.~A., \& {Breen}, S.~L. 2014, \mnras, 439, 2584, \dodoi{10.1093/mnras/stu116}

\bibitem[{{Walsh} {et~al.}(2003){Walsh}, {Macdonald}, {Alvey}, {Burton}, \& {Lee}}]{2003A&A...410..597W}
{Walsh}, A.~J., {Macdonald}, G.~H., {Alvey}, N.~D.~S., {Burton}, M.~G., \& {Lee}, J.~K. 2003, \aap, 410, 597, \dodoi{10.1051/0004-6361:20031191}

\bibitem[{{Walsh} {et~al.}(2014){Walsh}, {Purcell}, {Longmore}, {Breen}, {Green}, {Harvey-Smith}, {Jordan}, \& {Macpherson}}]{2014MNRAS.442.2240W}
{Walsh}, A.~J., {Purcell}, C.~R., {Longmore}, S.~N., {et~al.} 2014, \mnras, 442, 2240, \dodoi{10.1093/mnras/stu989}

\bibitem[{{Xu} {et~al.}(2008){Xu}, {Li}, {Hachisuka}, {Pandian}, {Menten}, \& {Henkel}}]{Xu2008}
{Xu}, Y., {Li}, J.~J., {Hachisuka}, K., {et~al.} 2008, \aap, 485, 729, \dodoi{10.1051/0004-6361:200809472}

\bibitem[{{Xue} {et~al.}(2019){Xue}, {Remijan}, {Burkhardt}, \& {Herbst}}]{2019ApJ...871..112X}
{Xue}, C., {Remijan}, A.~J., {Burkhardt}, A.~M., \& {Herbst}, E. 2019, \apj, 871, 112, \dodoi{10.3847/1538-4357/aaf738}

\bibitem[{{Yan} {et~al.}(2022){Yan}, {Henkel}, {Menten}, {Gong}, {Nguyen}, {Ott}, {Ginsburg}, {Wilson}, {Brunthaler}, {Belloche}, {Zhang}, {Budaiev}, \& {Jeff}}]{2022A&A...666L..15Y}
{Yan}, Y.~T., {Henkel}, C., {Menten}, K.~M., {et~al.} 2022, \aap, 666, L15, \dodoi{10.1051/0004-6361/202245024}

\bibitem[{{Yang} {et~al.}(2023){Yang}, {Gong}, {Menten}, {Urquhart}, {Henkel}, {Wyrowski}, {Csengeri}, {Ellingsen}, {Bemis}, \& {Jang}}]{Yang2023}
{Yang}, W., {Gong}, Y., {Menten}, K.~M., {et~al.} 2023, arXiv e-prints, arXiv:2305.04264, \dodoi{10.48550/arXiv.2305.04264}

\end{thebibliography}

\newpage

\appendix
\section{Infrared Observations on Possible Associated Point Sources} \label{apx:IR}
\begin{figure*}[h]
    \centering
    \includegraphics[width=\textwidth]{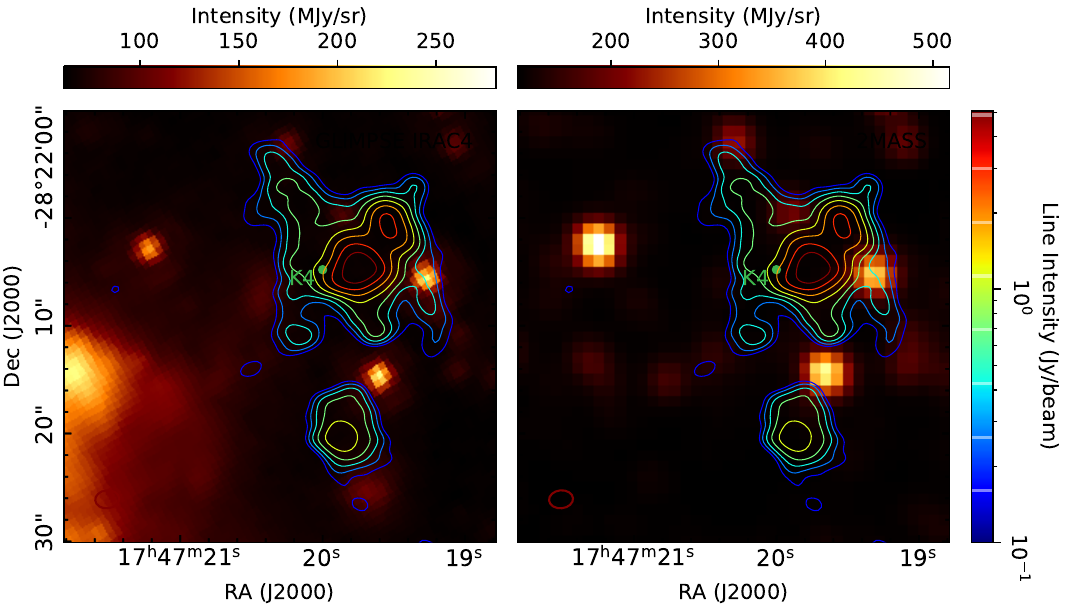}
    \caption{Infrared sources towards Sgr B2(N) reported by the GLIMPSE IRC 4 Band at $8.0\mu m$ (Left) and the 2MSAA Ks Band at $2.2\mu m$ (Right). The 84 GHz CH3OH maser emission at $58.5\,\mathrm{km\,s^{-1}}$ is shown with contours, as in Figure~\ref{fig:ch3oh-84GHz-map}($r$).
    }
    \label{fig:ir-sources}
\end{figure*}

\restartappendixnumbering
We searched the point source catalog from both the 2MASS and the GLIMPSE program with the coordinate of MS1 ($\alpha_\mathrm{J2000}=\hms{17}{47}{19.73}$, $\delta_\mathrm{J2000}=\dms{-28}{22}{04.58}$). As shown in Figure~\ref{fig:ir-sources}, both 2MASS and GLIMPSE observations show that the nearest infrared source is located at $\alpha_\mathrm{J2000}=\hms{17}{47}{19.34}$, $\delta_\mathrm{J2000}=\dms{-28}{22}{05.70}$, $\sim5.2\arcsec$ west of the maser spot. In addition, GLIMPSE detected another weak point source located toward $\alpha_\mathrm{J2000}=\hms{17}{47}{19.60}$, $\delta_\mathrm{J2000}=\dms{-28}{22}{06.82}$, $\sim2.8\arcsec$ southwest of the maser spot.

\section{MCMC Fitting Details \label{apx:mcmc}}
\begin{figure}
    \centering
    \includegraphics[width=\textwidth]{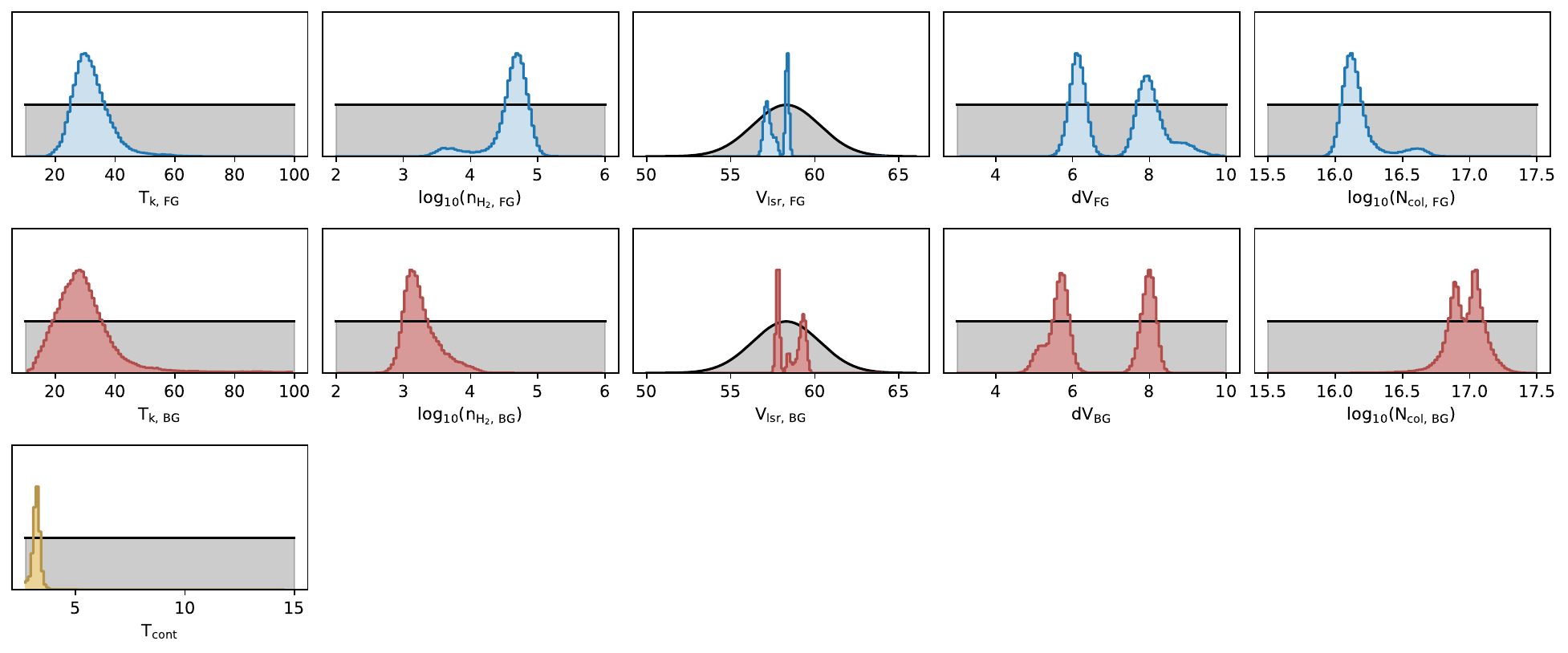}
    \caption{\label{fig:ch3oh-prior-posterior-II} Prior and posterior distributions for the MCMC fit of \ce{CH3OH} $E$-type transitions toward MS1. The prior distribution of each parameter is shown in gray. The bimodal posterior distributions for the foreground component are shown in blue, whereas those for the background component are shown in red. The posterior of $T_{cont}$ is shown in yellow.}
\end{figure}

\restartappendixnumbering
\paragraph{$E$-type \ce{CH3OH}} There are 11 free parameters in total to be adjusted in the MCMC analyses for $E$-type \ce{CH3OH}. For the spectra extracted toward MS1, the adopted prior and resultant posterior probability distributions for each parameter are depicted in Figure~\ref{fig:ch3oh-prior-posterior-II}. The priors for $V_\mathrm{lsr}$ of both components are chosen to be a Gaussian distribution centered at $58.3\,\mathrm{km\,s^{-1}}$ with a standard deviation of $2\,\mathrm{km\,s^{-1}}$. Since the continuum is weak as evident by the observations, we opted a uniform prior distribution in the range of 2.725 $-$ 15 K for $T_\mathrm{cont}$. Meanwhile, all the remaining parameters have uninformative uniformly distributed priors. We found that the convergence rate of the \texttt{RADEX} algorithm slows down tremendously at high $N_\mathrm{col} > 10^{17}\,\mathrm{cm^{-2}}$. Therefore, we capped $N_\mathrm{col}$ at $10^{17.5}\,\mathrm{cm^{-2}}$ and used 50,000 samples per walker (five times the value used for MS2). At high $N_\mathrm{col}$, we observed deviations between the results obtained from our modified non-LTE algorithm and those from \texttt{RADEX} due to results not being converged in the latter.

\begin{figure*}
    \centering
    \includegraphics[width=\textwidth]{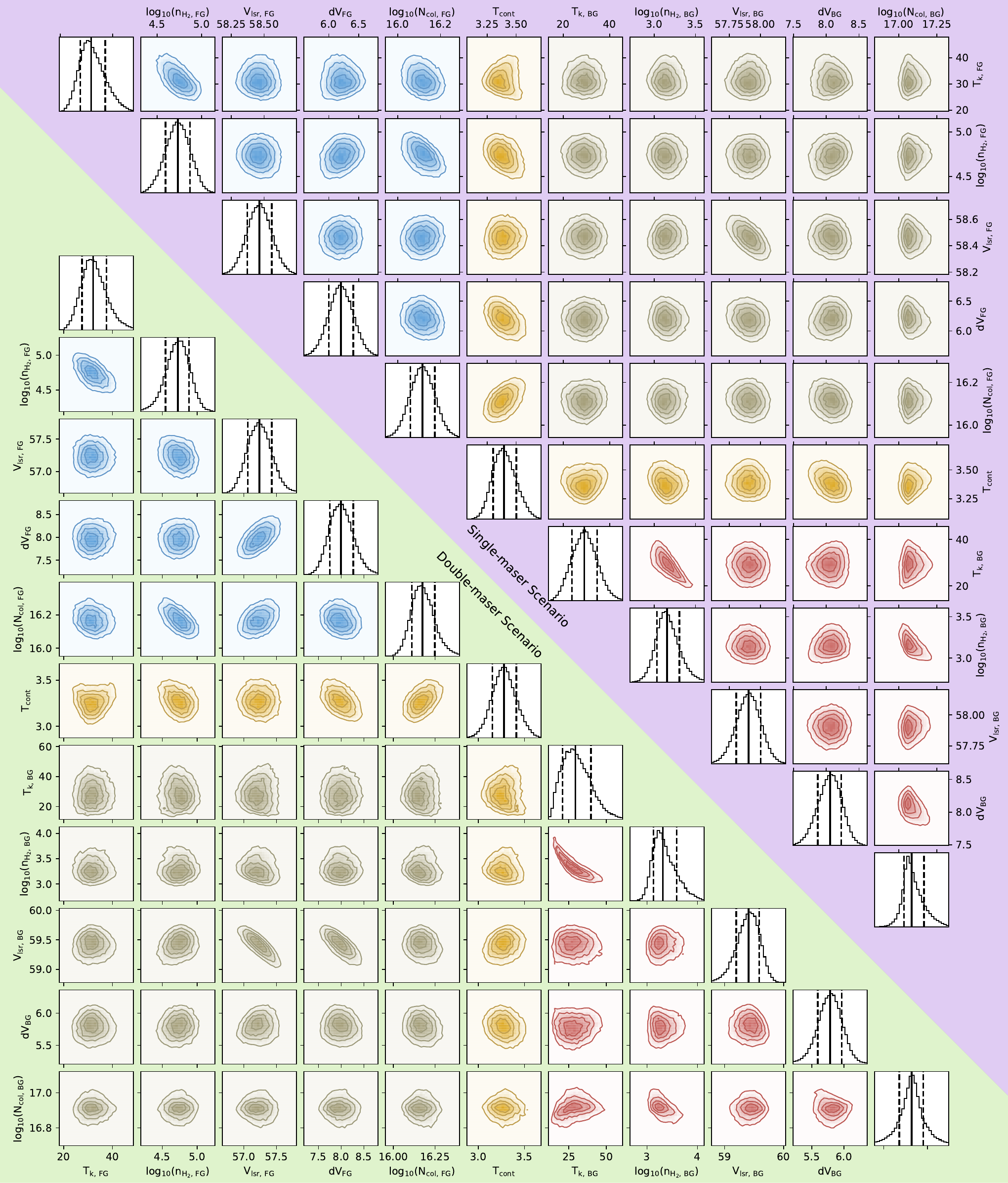}
    \caption{\label{fig:MeOH_triangle_58kms} Parameter covariances and marginalized posterior distributions for the MCMC fit of \ce{CH3OH} $E$-type transitions toward MS1. The 16$^{th}$, 50$^{th}$, and 84$^{th}$ confidence intervals (corresponding to $\pm$1 sigma for a Gaussian posterior distribution) are shown as vertical lines. The distribution among the parameters for the foreground component is shown in blue and that for the background component is shown in red. The distribution among the parameters between the two components is shown in brown. The distribution between the background continuum temperature and other parameters is shown in yellow. To better visualize the bimodal posteriors, we isolated the two peaks and plotted separate posteriors for each, with the double-maser and single-maser scenarios highlighted by green and purple backgrounds respectively, as per the color scheme in Figure~\ref{fig:ch3oh-58kms-spectra}.}
\end{figure*}

A corner plot of the parameter covariances and their distribution for the MS1 MCMC fit is shown in Figure~\ref{fig:MeOH_triangle_58kms}. As shown in Figure~\ref{fig:ch3oh-prior-posterior-II}, we found that the optimal set of parameters converge to two peaks, each corresponding to one of the two scenarios (double-maser and single-maser). In particular, the peak at lower velocity of the foreground component, corresponding to the higher velocity peak of the background component, corresponds to the double-maser scenario, while the other peak corresponds to the single-maser scenario. To better visualize the posteriors, we isolated the two peaks using $V_\mathrm{lsr} = 58\,\mathrm{km\,s^{-1}}$ as the criterion and plotted the posteriors separately in Figure~\ref{fig:MeOH_triangle_58kms}. The corner plots with green and purple backgrounds correspond to the double-maser and single-maser scenarios, respectively.

In each corner, the first five parameters belong to the foreground component, while the last five are those of the background component. The 2D posterior distribution among parameters for the foreground component is shown in blue, while that for the background component is shown in red. The cross terms between the two components are shown in brown. The middle parameter is the background continuum temperature and the distribution between continuum temperature and other parameters is shown in yellow.

\begin{figure*}
    \centering
    \includegraphics[width=\textwidth]{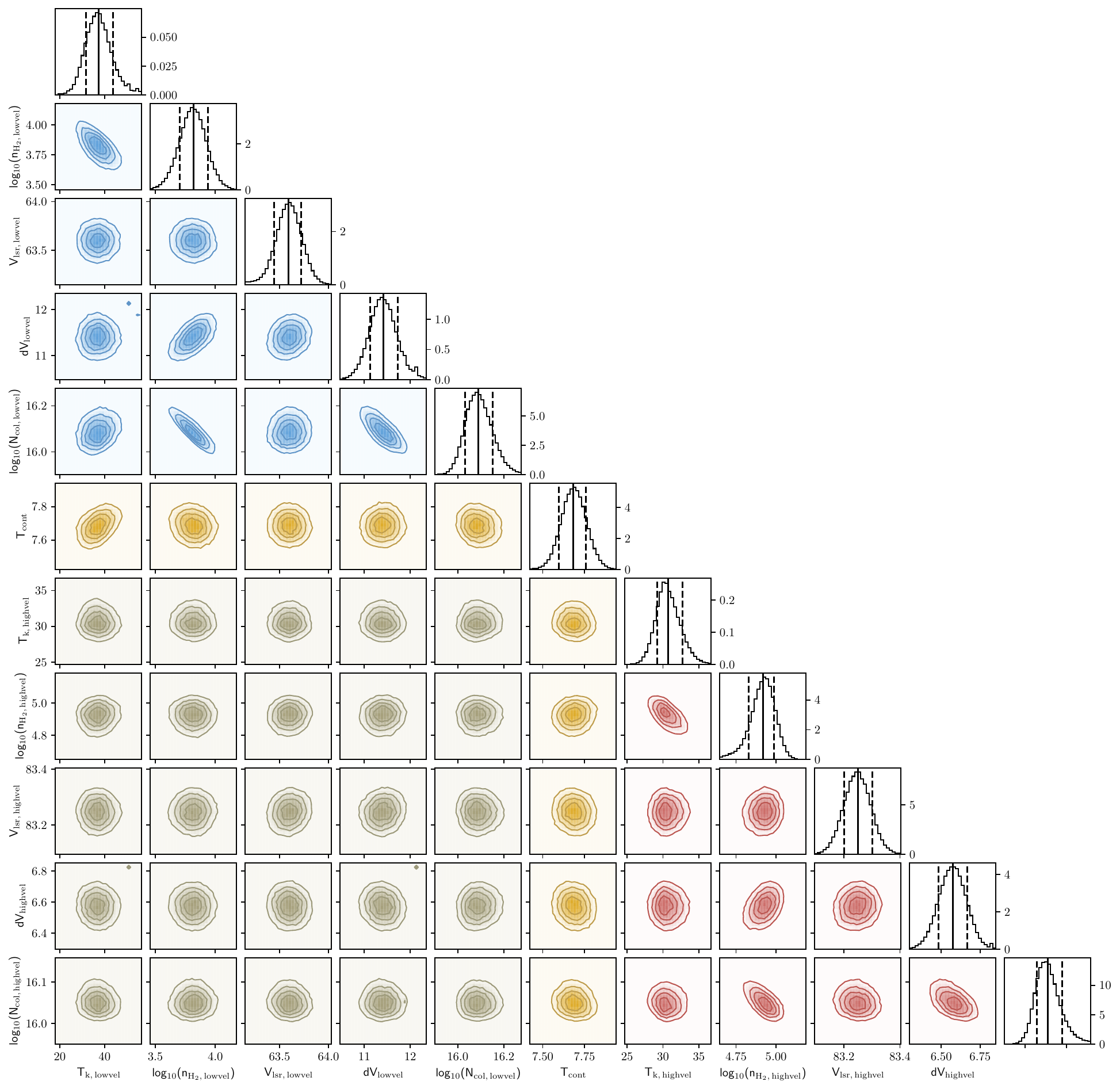}
    \caption{\label{fig:MeOH_triangle_83kms}
     Similar to Figure~\ref{fig:MeOH_triangle_58kms}, parameter covariances and marginalized posterior distributions for the MCMC fit of \ce{CH3OH} $E$-type transitions toward MS2.} The distribution among the parameters for the low-velocity (64-$\mathrm{km\,s^{-1}}$) component are shown in blue and that for the high-velocity (83-$\mathrm{km\,s^{-1}}$) component is shown in red. The distribution among the parameters between the two components are shown in brown. The distribution between background continuum temperature and other parameters is shown in yellow.
\end{figure*}

A similar MCMC analysis to that for the spectra extracted toward MS1 was carried out for the spectra extracted toward MS2. In contrast to Figure~\ref{fig:ch3oh-prior-posterior-II}, the optimal set of parameters of MS2 converges to a single peak as shown in Figure~\ref{fig:ch3oh-prior-posterior-I}. The corresponding corner plot of the parameter covariances and their distribution for the MS2 MCMC fit are shown in Figure~\ref{fig:MeOH_triangle_83kms}. Similar to Figure~\ref{fig:MeOH_triangle_58kms}, the parameters of the low-velocity component are shown in blue while those of the high-velocity component are shown in red. We found that the MCMC analyses yield the consistent results, regardless of whether the foreground or background velocity component is assigned as the high or low velocity along the LOS. Furthermore, the brown circular distribution, the cross terms between the two components, reveals that the two components are completely independent. This independence arises because the two components fit distinct portions of the spectrum because of their disparate central velocities compared with the line widths. In contrast, the strong degeneracy of the \ce{H2} density and \ce{CH3OH} column density is revealed for each individual velocity component.

\begin{figure*}
    \centering
    \includegraphics[width=0.7\textwidth]{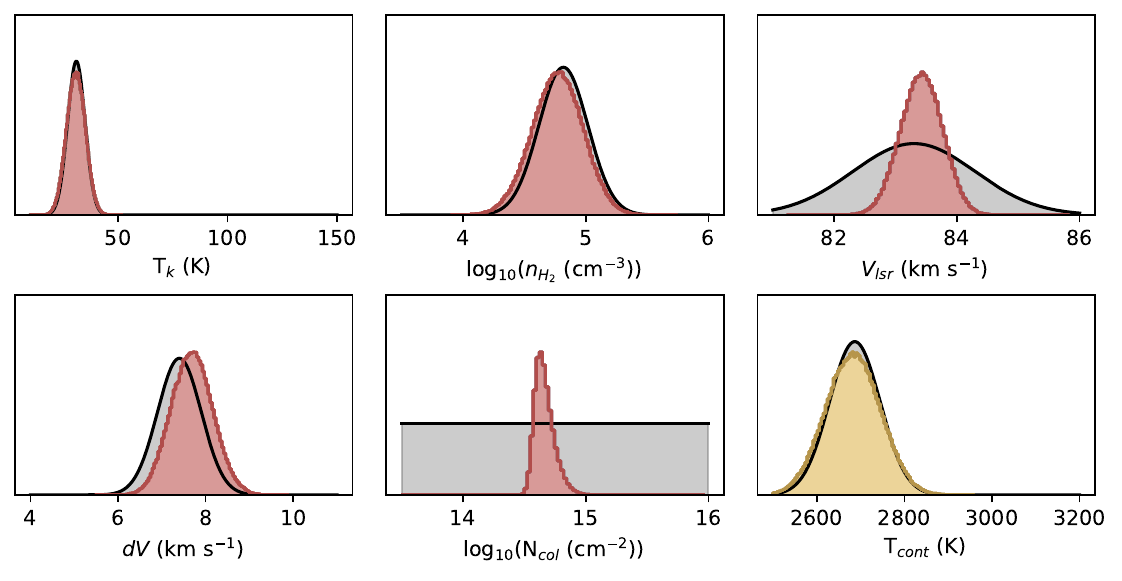}
    \caption{\label{fig:ch2nh-prior-posterior} Prior and posterior distributions for the MCMC fit of the \ce{CH2NH} transitions toward MS2. The prior distribution for each parameter is shown in grey, while the priors on $T_k$ and $n_{\ce{H2}}$ are from the MCMC posteriors of \ce{CH3OH} in MS2 (Figure~\ref{fig:ch3oh-prior-posterior-I}). The posterior distributions for the physical characteristics are shown in red while the posterior for $T_{cont}$ is shown in yellow.}
\end{figure*}

\begin{figure*}
    \centering
    \includegraphics[width=0.7\textwidth]{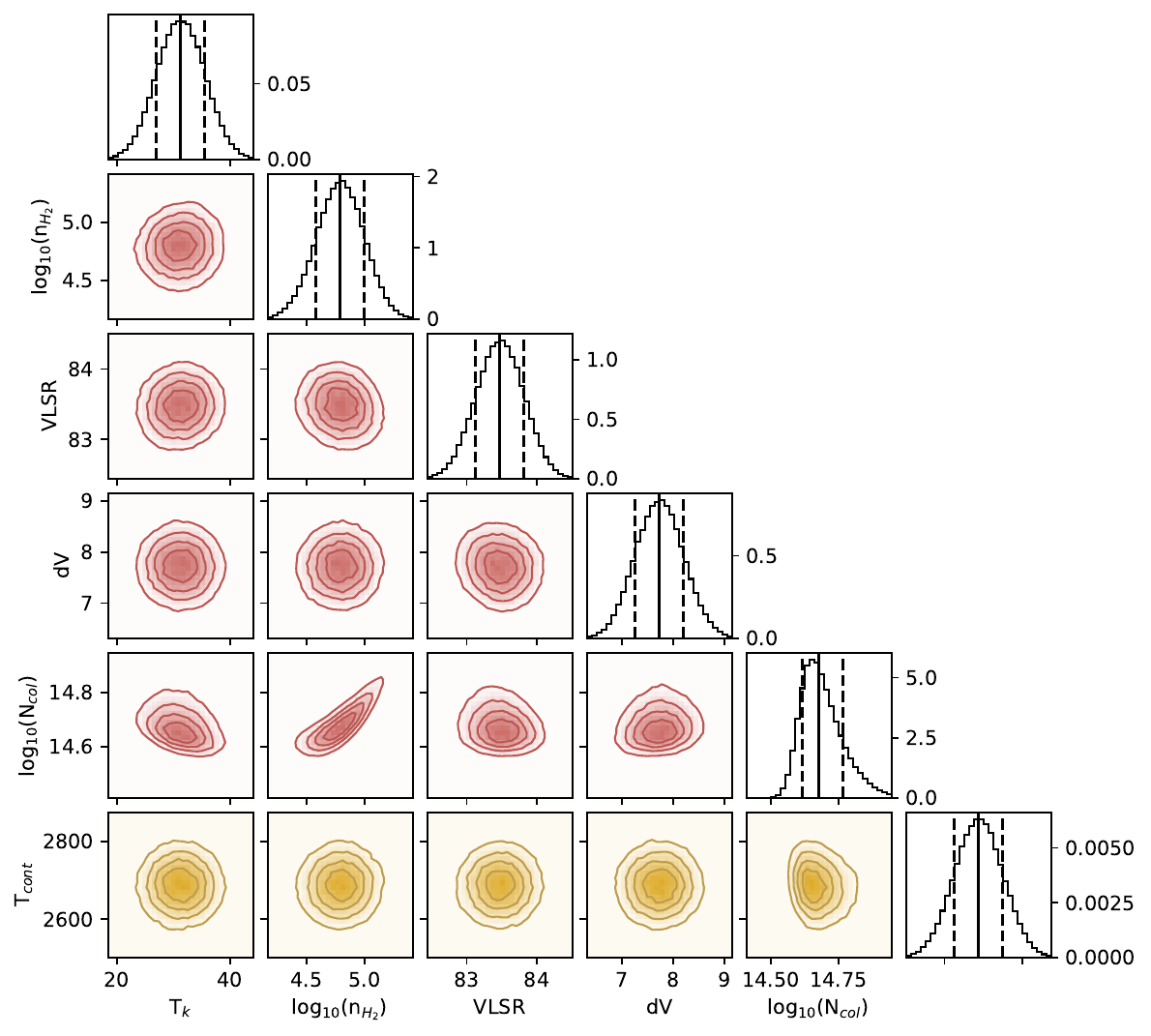}
    \caption{\label{fig:CH2NH_triangle_83kms} Parameter covariances and marginalized posterior distributions for the MCMC fit of the \ce{CH2NH} transitions. The 16$^{th}$, 50$^{th}$, and 84$^{th}$ confidence intervals (corresponding to $\pm$1 sigma for a Gaussian posterior distribution) are shown as vertical lines. The distribution among the parameters for the masing gas are shown in red and the distribution between background continuum temperature and other parameters is shown in yellow. }
\end{figure*}

\paragraph{\ce{CH2NH}} An MCMC analysis with a one-clump model was carried out for the \ce{CH2NH} spectra. The prior and posterior distributions are shown in Figure~\ref{fig:ch2nh-prior-posterior}, while a corner plot of the parameter covariances for the \ce{CH2NH} MCMC fit is shown in Figure~\ref{fig:CH2NH_triangle_83kms}.

\section{Collisional Rate Coefficient Calculations for \texorpdfstring{\ce{CH2NH}}{CH2NH}} \label{apx:ch2nh-coll}
A first set of collisional data for \ce{CH2NH} + p-\ce{H2}($j_2=0$) was
published by \citet{2018JPCL....9.3199F} for applications to cold gas.  The calculations of \citet{2018JPCL....9.3199F} were performed for rotational transitions among the first 15 rotational levels of \ce{CH2NH}, i.e. up to the state $2_{2, 0}$ ($j_1=2, K_a=2, K_c=0$) with an energy of 28.3~cm$^{-1}$. The full close-coupling method was used to generate rate coefficients at temperatures $5 \leq T_k \leq 30$~K and hyperfine-resolved rate coefficients were derived from the rotational
rate coefficients using the statistical approximation, see \citet{2018JPCL....9.3199F} for full details. 

In this work, we have extended the calculations of \citet{2018JPCL....9.3199F} to higher levels and temperatures and we have thus considered o-\ce{H2}($j_2=1$) as a second collider. Because close-coupling calculations are computationally very demanding when the temperature increases, we had recourse to the coupled-states approximation. As in \citet{2018JPCL....9.3199F}, all calculations were performed with the OpenMP extension \footnote{\footnotesize{http://ipag.osug.fr/$\sim$faurea/molscat/index.html}} of the version 14 of the \texttt{MOLSCAT} code \citep{molscat}, using the same propagation parameters and rotational basis for \ce{CH2NH} ($j_1\leq12$) and p-\ce{H2} ($j_2=0, 2$). The basis for o-\ce{H2} was
restricted to $j_2=1$ and the \texttt{MOLSCAT} parameter \texttt{EMAX} (used to limit basis functions to \ce{CH2NH} plus \ce{H2} energies less than \texttt{EMAX}) was set to 1200~cm$^{-1}$ for p-\ce{H2} and 1318.64~cm$^{-1}$ for o-\ce{H2}. The coupled-states calculations were performed for total energies between 2 and 800 cm$^{-1}$ for p-\ce{H2} and between 120 and 918~cm$^{-1}$ for o-\ce{H2}. Rate coefficients were obtained for all rotational transitions among the first 49 levels, i.e. up to the state $8_{2, 6}$ at 99.5~cm$^{-1}$, and for kinetic temperatures in the range 10 -- 150~K. Hyperfine-resolved rate coefficients were derived using the statistical approximation, as in \citet{2018JPCL....9.3199F}. Finally, the collisional dataset was combined with the energy levels and radiative rates taken from the CDMS catalog \citep{2005JMoSt.742..215M} in order to provide an EMAA-formatted datafile.

\end{document}